\documentclass[%
 aip,
 amsmath,amssymb,
 reprint,%
]{revtex4-1}

\usepackage{graphicx}
\usepackage{bm}

\usepackage[utf8]{inputenc}
\usepackage[T1]{fontenc}
\usepackage{mathptmx}
\usepackage{etoolbox}
\usepackage{color}

\begin{document}
\preprint{AIP/123-QED}
\title{Spin and spin current---From fundamentals to recent progress}
%
\author{Sadamichi Maekawa}
\email{sadamichi.maekawa@riken.jp}
\affiliation{Center for Emergent Matter Science (CEMS), RIKEN, Wako 351-0198, Japan}
\affiliation{Advanced Science Research Center, Japan Atomic Energy Agency, Tokai 319-1195, Japan}
\affiliation{Kavli Institute for Theoretical Sciences, University of Chinese Academy of Sciences, Beijing, 100190, China}
\author{Takashi Kikkawa}
\affiliation{Department of Applied Physics, The University of Tokyo, Tokyo 113-8656, Japan}
\author{Hiroyuki Chudo}
\affiliation{Advanced Science Research Center, Japan Atomic Energy Agency, Tokai 319-1195, Japan}
\author{Jun'ichi Ieda}
\affiliation{Advanced Science Research Center, Japan Atomic Energy Agency, Tokai 319-1195, Japan}
\author{Eiji Saitoh}
\email{eizi@ap.t.u-tokyo.ac.jp}
\affiliation{Advanced Science Research Center, Japan Atomic Energy Agency, Tokai 319-1195, Japan}
\affiliation{Department of Applied Physics, The University of Tokyo, Tokyo 113-8656, Japan}
\affiliation{Institute for AI and Beyond, The University of Tokyo, Tokyo 113-8656, Japan}
\affiliation{WPI Advanced Institute for Materials Research, Tohoku University, Sendai 980-8577, Japan}
\date{\today}

\begin{abstract}
Along with the progress of spin science and spintronics research, the flow of electron spins, \emph{i.e.}, spin current, has attracted interest. New phenomena and electronic states were explained in succession using the concept of spin current. 
Moreover, as many of the conventionally known spintronics phenomena became well organized based on spin current, it has rapidly been recognized as an essential concept in a wide range of condensed matter physics. 
In this article, we focus on recent developments in the physics of spin, spin current, and their related phenomena, where the conversion between spin angular momentum and different forms of angular momentum plays an essential role. Starting with an introduction to spin current, we first discuss the recent progress in spintronic phenomena driven by spin-exchange coupling: spin pumping, topological Hall torque, and emergent inductor. We, then, extend our discussion to the interaction/interconversion of spins with heat, lattice vibrations, and charge current and address recent progress and perspectives on the spin Seebeck and Peltier effects. Next, we review the interaction between mechanical motion and electron/nuclear spins and argue the difference between the Barnett field and rotational Doppler effect. We show that the Barnett effect reveals the angular momentum compensation temperature, at which the net angular momentum is quenched in ferrimagnets. 
\end{abstract}
\maketitle

%
%
\section{Introduction}

Spin is an intrinsic property of an electron, which is associated with angular momentum, $ \hbar / 2 $, and magnetic moment, $\mu_{\rm e} = e \hbar / 2 m_ \mathrm{ e } c $, where $ e $, $ \hbar $, $ m_ \mathrm{ e } $, and $ c $ represent the electron charge, Planck constant divided by $ 2 \pi $, electron mass, and light velocity, respectively.  The spins interact with each other and align when they are in a ferromagnet.  Another intrinsic property of an electron is charge, $ e $, which is utilized in electronics.  Oxford dictionary defines electronics as ``the branch of physics and technology concerned with the design of circuits using transistors and microchips, and with the behavior and movement of electrons in a semiconductor, conductor, vacuum, or gas.''  Here, ``movement of electrons'' means ``the movement of electron charges.''  Electron spins can also flow in a material, independently of their charges.  This is called \emph{ spin current}\cite{Spin-current-text,spincurrent_JPSJ}, originally predicted by D'yakonov and Perel' \cite{Spin-current-Dyakonov1,Spin-current-Dyakonov2} in 1971.  Since the beginning of this century, special attention has been paid to spin current and \emph { spin accumulation } in semiconductors, metals, and even insulators, and to how to utilize them together with electric charge current.  This field of research and applications is named \emph {spintronics}, in contrast with electronics.

  The flows of spin and charge of electrons can couple each other due to spin-orbit coupling (SOC), as well as electromagnetic interaction.  As a result, spin current and electric current may convert each other.
The interconversion between the currents is called the \emph { spin Hall effect } (SHE) and \emph { inverse spin Hall effect} (ISHE)~\cite{Hirsch1999PRL,Saitoh2006APL,Azevedo2005JAP,Valenzuela2006Nature,Costache2006PRL,Kimura2007PRL}, and is one of the key principles in spintronics.
Concurrently, the discovery of \emph{spin-transfer torque} (STT)~\cite{Slonczewski_1989,Slonczewski1996JMMM,Berger_1996,Myers_1999} brought about significant advances in applications of spin current for magnetic memory technology.
Additionally, it was shown recently that spin and charge interact due to the geometrical effect called \emph { spin Berry phase}~\cite{Stern_1992, Nagaosa_2012, Nagaosa_2013}.  The spin Berry phase also gives rise to a variety of the interconversion of spin and charge.  This is also important in spintronics.

  It is known that nuclei, such as protons and neutrons, have spin angular momenta which are of the same order of magnitude of electron spins, although their magnetic moments, $\mu_{\rm N}$, are smaller by three orders of magnitude than that of an electron, $\mu_{\rm e}$ ($\mu_{\rm N}/\mu_{\rm e} = m_ \mathrm{ e } / m_ \mathrm{ N } \sim 10^{-3}$ with $ m_ \mathrm{ N } $ being the mass of a proton). Therefore, considering angular momentum conservation, nuclear spins may contribute to a variety of physical properties in a material.  The conservation may also include the angular momenta of mechanical motion and vortices of the flow of the constituents of a material.  This suggests that spins of electrons couple to the mechanical motions of a material.

  The purpose of this article is to discuss various aspects of materials based on angular momentum conservation law, including not only electron spins, but nuclear spins and mechanical motion. In the second section, the concepts of spin current, spin accumulation, and other quantities important in spintronics are explained.  In the third section, the interaction and interconversion of spin current and charge current are discussed.  The generalization of Faraday's law of electromagnetic induction to spintronics is presented\cite{Barnes_2007}.  In the fourth section, the interaction of spins with heat, lattice vibrations, and electric current is introduced. Recent progress in and perspectives on the spin Seebeck and Peltier effects, which are two representative spin-heat coupling phenomena, are overviewed.  The behavior of nuclear spins is examined in connection with that of electron spins.  In the fifth section, the interaction between mechanical motion and electron/nuclear spins is examined.  The interaction is called the Einstein-de Hass effect\cite{Einstein1915} whose inverse is the Barnett effect\cite{Barnett1915}.  These interactions were discovered in 1915, but have not received much attention until recently.  The topics in this section present recent experiments and perspectives of these effects. The last section summarizes the present article. 
  

%
%
\section{Concept of Spin Current}
\subsection{Conduction-electron spin current}
\begin{figure}[tbh]
\begin{center}
\includegraphics[width=8.5cm]{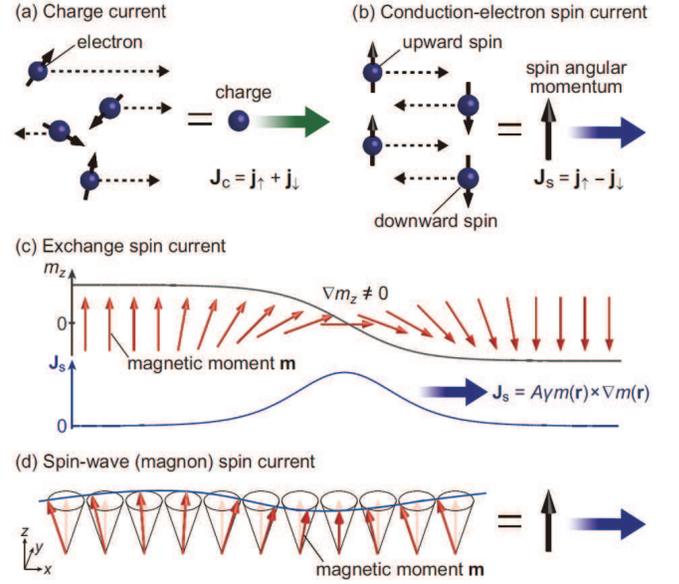}
\caption{
Illustrations of the (a) charge current, (b) conduction-electron spin current, (c) exchange spin current, and (d) spin-wave (magnon) spin current.
}\label{fig:Spin-current}
\end{center}
\end{figure}

When an electric field is applied to a conductor, there is a slight difference in electrons quantities having a velocity component parallel and antiparallel to the electric field. Due to this slight imbalance in the electron velocity distribution, a group of electrons flows in a certain direction, which is an electric (charge) current [Fig. \ref{fig:Spin-current}(a)]. Then, what would happen, when the electrons traveling in the opposite direction tend to have spins in opposite directions  [Fig. \ref{fig:Spin-current}(b)]? When the number of electrons traveling in both directions is the same, the charge flows cancel out each other and the net current is zero. By contrast, the spin flow remains uncanceled. This imbalance increases the upward spin on the right edge of the sample and reduces the upward spin at the left end. Thus, there is an upward spin flow from left to right. (This can be rephrased as a downward spin flow from right to left.) In this way, by introducing such imbalance, it is possible to create only the spin flow with zero electric current. This is a spin current carried by conduction electrons, which is called conduction-electron spin current [Fig. \ref{fig:Spin-current}(b)]. The spin $ z $ component of the spin current is expressed as
\begin{equation}
\mathbf{ J }_\mathrm{ s } = \mathbf{ j }_\uparrow - \mathbf{ j }_\downarrow . 
\end{equation}
The current $ \mathbf{ j }_{\uparrow ( \downarrow ) } $ represents a particle flow carrying the upward (downward) spin. By contrast, a charge current is described as
\begin{equation}
\mathbf{ J }_\mathrm{ c } = \mathbf{ j }_\uparrow + \mathbf{ j }_\downarrow . 
\end{equation}
Here, we introduce the spin dependent chemical potentials, $ \mu_\uparrow $ and $\mu_\downarrow $, which satisfy
\begin{equation}
\mathbf{ j }_{ \uparrow ( \downarrow ) } = D \nabla \mu_{ \uparrow ( \downarrow ) } ,
\end{equation}
where $ D $ is the electron diffusion constant. 
By using $ \mu_{ \uparrow ( \downarrow ) } $, the spin current, $\mathbf{J}_\mathrm{s}$, can be written as
\begin{equation}
\mathbf{ J }_\mathrm{ s } = D \nabla \left( \mu_\uparrow - \mu_\downarrow \right)  . 
\end{equation}
This shows that $ ( \mu_\uparrow - \mu_\downarrow ) $ acts as a driving force for conduction-electron spin current. 
In fact, as the spin current is not a conservative flow, it is difficult to transport the spin current over long distances. This is different from a charge current. As a charge current is a conservative current, if a current is passed through one end of a metal wire, the same current can be taken out from the other end, unless the wire is charged because of charge conservation law. By contrast, there is no such strong conservation law for spin, and there is a certain probability that the spin direction will naturally change to disorder. As a result, the conduction-electron spin current disappears after flowing over a certain distance, which is called the spin diffusion length. On a scale sufficiently shorter than the spin diffusion length, the spin current can be regarded as an approximate conservative current and can be expressed in the same way as the charge current.  

Note that, in a strict sense, the physics of spin current cannot be constructed in exactly the same manner as electric current. In fact, as described below, the definition of spin current is not uniquely determined because conservation law does not hold exactly for spin. This makes it challenging to provide an accurate definition for spin current. Nevertheless, the introduction of spin current has contributed to new physical phenomena discovery. The discovery greatly advances the understanding of spin current. 
The concept of spin current has been a powerful guiding principle for the development of modern (condensed matter) physics.  

Let us define spin current under the assumption that the spin is a good quantum number and the spin angular momentum is fully conserved. In general, the magnetic moment of electrons carries the angular momentum, where the ratio is given by the gyromagnetic ratio, $ \gamma $ ($< 0$). Then, the angular momentum conservation is written as
\begin{equation}
\frac{ \partial }{ \partial t }  \mathbf{ M }  ( \mathbf{ r }, t ) =  - \gamma\ \mathrm{div} \mathbf{ J }_\mathrm{ s }  ( \mathbf{ r }, t ) , 
\label{eq.continuity}
\end{equation}
where $ \mathbf{ M } $, $ \mathbf{ r } $, and $t$ denote the local magnetization (magnetic-moment density), the position of the electron, and time, respectively. 
This equation shows that when $ \mathbf{ J }_\mathrm{ s }  $ changes spatially in a magnet, the magnetization experiences torque (spin transfer torque) and it starts rotating. Equation (\ref{eq.continuity}) can be used as an acceptable approximation on a microscale, where the conservation of spin angular momentum is considered to be approximately valid. However, we need to keep in mind that, in general, the spin angular-momentum conservation law does not hold to some extent. 

\subsection{ Spin relaxation }
Next, let us consider that spin is not conserved. The simplest phenomenological expression of spin relaxation is 
\begin{equation}
 \mathbf{ T }  = - \frac{ \mathbf{ M }  ( \mathbf{ r } ) -  \mathbf{ M  }_0  }{ \tau }  , 
\end{equation}
where $ \tau $  is the spin relaxation time. If this relaxation term is incorporated, Eq.~(\ref{eq.continuity}) becomes
\begin{equation}
\frac{ \partial }{ \partial t }  \mathbf{ M }  ( \mathbf{ r } ) =  - \gamma\ \mathrm{div}  \mathbf{ J }_\mathrm{ s }  +  \mathbf{ T } . 
\end{equation}
The mechanism of spin relaxation in materials is quite complicated. To date, several models for spin relaxation mechanisms have been proposed. One such model is the mechanism proposed by D'yakonov and Perel~\cite{Dyakonov-Perel}. In this model, electron spins undergo precessional rotational motion due to the spin-orbit interaction, and the direction of precession changes as the direction of motion changes from electron scattering during orbital motion. In this model, a longer electron scattering time allows the spins to rotate by a larger angle, resulting in a stronger spin relaxation. Using an approximation with a high electron scattering frequency, the spin relaxation time is given by
\begin{equation}
\frac{ 1 }{ \tau } \sim \omega ^2 \tau _ p,
\end{equation}
where $ \omega $ is the angular frequency of spin precession and $\tau _ p $ is the electron momentum relaxation time. When conduction electrons are scattered, their spin can flip with the help of the spin-orbit interaction. This is called the Elliott---Yafet mechanism~\cite{Elliott1954PhysRev,Yafet1963SolidStatePhys}. In this mechanism, spin relaxation strength is proportional to electron scattering probability, $  1 / \tau  \propto  1 / \tau _ p $.

The existence of spin relaxation obscures the definition of spin current, as discussed above. The conserved current is uniquely determined by the continuity equation. However, the spin current has a relaxation term, and the continuity equation [Eq. (\ref{eq.continuity})] does not remain consistent, which makes it impossible to accurately define the spin current. Any term extracted from $ \mathbf{ T }  $ can be added to spin current, $  \mathbf{ J }_\mathrm{ s } $. In such cases, the definition of spin current must be considered on a case-by-case basis, depending on the object.  

\subsection{ Exchange spin current }
An equation of motion for magnetization is given by
\begin{equation}
\frac{ d }{ d t }  \mathbf{ m } =  - \gamma\ \mathbf{ m } \times  \mathbf{ H }_\mathrm{ eff } . 
\label{eq.LL}
\end{equation}
Here, $ \mathbf{ m } $ denotes the magnetic moment per volume and $ \mathbf{ H }_\mathrm{ eff }  $ represents the effective field acting on $ \mathbf{ m } $, 
\begin{equation}
\mathbf{ H }_\mathrm{ eff }  = \frac{ \delta E ( \mathbf{ m } ) }{ \delta \mathbf{ m } } ,
\end{equation}
where $ E ( \mathbf{ m } ) $ is the total magnetic energy. 
Let us consider the effective magnetic field due to the ferromagnetic (Heisenberg) exchange interaction,
\begin{equation}
\mathcal{ H }  = -2 J \sum _{ \langle i, j \rangle } \mathbf{ S }_i  \cdot \mathbf{ S }_j ,
\end{equation}
where $\mathbf{ S }_i$ is the spin of the $i$-th lattice site,  $ J $  is called the exchange interaction constant, and the sum is taken over the nearest neighbor pairs. The effective field at the site, $i$, thereby reads as
\begin{equation}
\mathbf{ H }_{\mathrm{ eff }, i}  =  - 2 \frac{ J }{ \gamma } \sum _{ \langle i, j \rangle } \mathbf{ S }_j. \label{eq.effective_exchange_field}
\end{equation}
Substituting Eq. (\ref{eq.effective_exchange_field}) into the equation of motion [Eq. (\ref{eq.LL})] yields
\begin{equation}
\frac{ d }{ d t }  \mathbf{ m } =  2J \mathbf{ m } \times   \sum _{ \langle i, j \rangle } \mathbf{ S }_j.
\end{equation}
By using the continuum approximation,
\begin{equation}
 \mathbf{ S } ( \mathbf{ r } + \mathbf{ a } ) =   \mathbf{ S } ( \mathbf{ r } ) +  \frac{ \partial \mathbf{ S } ( \mathbf{ r } ) }{ \partial \mathbf{ r }  }  \cdot \mathbf{ a } 
 + \frac{ 1 }{ 2 } \frac{ \partial ^2 \mathbf{ S } ( \mathbf{ r } ) }{ \partial \mathbf{ r } ^2 }  \cdot \mathbf{ a } ^2 + \dots ,
\end{equation}
where $  \mathbf{ r }  $ is the position of spin and $ a = | \mathbf{ a } |$ is the lattice constant, the equation of motion [Eq. (\ref{eq.LL})] can be written as
\begin{equation}
\frac{ d }{ d t }  \mathbf{ m } ( \mathbf{ r } )  =  \frac{ 2J \mathbf{ m } a^2 }{ \gamma } \mathbf{ m } ( \mathbf{ r } ) \times \nabla ^2 \mathbf{ m } ( \mathbf{ r } ) .
\end{equation}
 Here, the formula, $ \mathbf{ m } \times \nabla ^2 \mathbf{ m }  = \mathrm{div} \left( \mathbf{ m } \times \nabla \mathbf{ m } \right) , $
yields
\begin{equation}
\frac{ d }{ d t }  \mathbf{ m } ( \mathbf{ r } )  =  - A \gamma \ \mathrm{div} \left( \mathbf{ m } \times \nabla \mathbf{ m } \right) , \label{eq.continuity_equation_in_FM}
\end{equation}
where $ A =  2 J a ^2 / \gamma ^2 $ is the spin stiffness constant. 
By defining the current, $ \mathbf{ J }_\mathrm{ s }$,  as 
\begin{equation}
\mathbf{ J }_\mathrm{ s }  =   A \gamma  \mathbf{ m } ( \mathbf{ r } ) \times \nabla \mathbf{ m } ( \mathbf{ r } ) , \label{eq.exchange_spin_current}
\end{equation}
it becomes clearer that the equation of motion [Eq. (\ref{eq.continuity_equation_in_FM})] has the same form as the continuity equation [Eq. (\ref{eq.continuity})] and captures the spin angular-momentum conservation law in ferromagnets. 
$\mathbf{ J }_\mathrm{ s } $ defined by Eq. (\ref{eq.exchange_spin_current}) is  interpreted as a spin current carried by the exchange interaction, and is thereby called an exchange spin current \cite{Spin-current-text} [Fig. \ref{fig:Spin-current}(c)].

\begin{figure*}[tbh]
\begin{center}
\includegraphics[width=14cm]{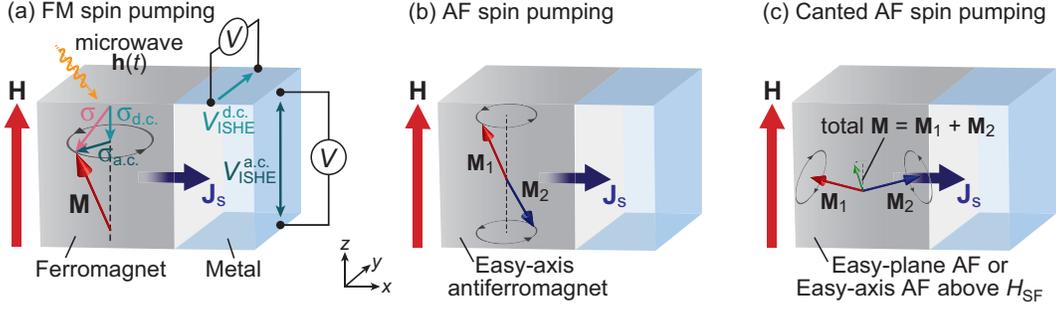}
\caption{Illustrations of the (a) ferromagnetic (FM), (b) antiferromagnetic (AF), and (c) canted AF spin pumping. 
When an eigenmode of the magnetization, ${\bf M}$, dynamics is resonantly excited by the applied radio-frequency (rf) field, ${\bf h}(t)$, a spin current, ${\bf J}_{\rm s}$, is emitted from the magnet into the attached metal via the interfacial spin-exchange coupling. As shown in (a), the nonequilibrium spin polarization, ${\mathbf \sigma}$, in spin pumping has d.c. (${\mathbf \sigma}_{\rm d.c.}$) and a.c. (${\mathbf \sigma}_{\rm a.c.}$) components: ${\mathbf \sigma} = {\mathbf \sigma}_{\rm d.c.} + {\mathbf \sigma}_{\rm a.c.}$ (d.c. = direct current, a.c. = alternating current). When ${\bf M}$ is oriented along the $z$ direction by the external magnetic field, ${\bf H}$, the d.c. polarization, ${\mathbf \sigma}_{\rm d.c.}$, appears parallel to ${\bf H}~||~{\bf {\hat z}}$, whereas the time-dependent ${\mathbf \sigma}_{\rm a.c.}$ rotates in the $xy$-plane. The d.c. (a.c.) ISHE voltage, $V_{\rm ISHE}^{\rm d.c.}$ ($V_{\rm ISHE}^{\rm a.c.}$), is generated in the metal along the $y$ ($z$) direction according to the relation ${\bf J}_{\rm s} \times {\mathbf \sigma}$, where ${\bf J}_{\rm s}~||~{\bf {\hat x}}$ \cite{Saitoh2006APL}.
}\label{fig:SP}
\end{center}
\end{figure*}

\subsection{ Spin-wave spin current } 
Next, consider the spin current carried by a spin wave [Fig. \ref{fig:Spin-current}(d)]~\cite{Kajiwara2010Nature}. As a simple example, we focus on exchange spin-waves propagating by an exchange interaction. We assume the small amplitude oscillation of the magnetization, $ \mathbf{ m } ( \mathbf{ r }, t )$, about the fixed direction, $ \mathbf{ m }_0 || \mathbf{ z }$, as
\begin{equation}
\mathbf{ m } ( \mathbf{ r }, t ) = \mathbf{ m }_0 + \mathbf{ m }_x ( \mathbf{ r }, t ) +\mathbf{ m }_y ( \mathbf{ r }, t ) \quad ( | \mathbf{ m }_0| \gg | \mathbf{ m }_{ x,y }| ),
\end{equation}
and we define
\begin{align}
\psi ( \mathbf{ r }, t )   & = m_x ( \mathbf{ r }, t ) + i m_y ( \mathbf{ r }, t ), \\
\psi ^ * ( \mathbf{ r }, t ) & = m_x ( \mathbf{ r }, t ) - i m_y ( \mathbf{ r }, t ). 
\end{align}
Using these expressions, the $ z $ component of the exchange spin current is rewritten as follows:
\begin{equation}
\mathbf{ j } ^ {M_z} \propto \psi ^* ( \mathbf{ r }, t ) \nabla \psi ( \mathbf{ r }, t ) -  \psi ( \mathbf{ r }, t ) \nabla \psi ^* ( \mathbf{ r }, t ) .
\end{equation}
By introducing magnons \cite{Spin-current-text,spincurrent_JPSJ}, the (exchange) spin current becomes 
\begin{equation}
\mathbf{ j } ^ {M_z} = \hbar \sum _ {\mathbf{ k } }\mathbf{ v }_{\mathbf{ k } } n _ { \mathbf{ k } }.
\end{equation}
Here, $ \mathbf { v }_{ \mathbf{ k } } $ is the group velocity of the spin wave, $ n_{ \mathbf{ k } } $ is the distribution function of the magnons, and $ \mathbf{ k } $ is the wave vector of spin waves. This formula means that when the magnon population is asymmetric between $ \mathbf{ k } $ and $ - \mathbf{ k } $, spin waves carry spin current.

%
%
\section{Spin-exchange coupling}

The interaction of spin current and magnetization dynamics has been a central interest in spintronics for a long time\cite{Spin-current-text,Hirohata_2020}, and it is continuously being developed for better and lower power consumption device applications\cite{Dieny_2020}. S-d exchange coupling is a key factor that controls spin-dependent transport and magnetization dynamics in ferromagnetic conductors 
and ferromagnetic/nonmagnetic interfaces.
This is due to the transfer of angular momentum and energy between conduction spin and magnetization texture. These characteristic properties are the basis for spin pumping \cite{Mizukami2002PRB,Tserkovnyak2002PRL,Ando2011JAP,Spin-current-text}, spin-transfer torque (STT) \cite{Slonczewski_1989,Slonczewski1996JMMM,Berger_1996,Myers_1999}, and spinmotive force (SMF)\cite{Berger_1986,Volovik_1987,Barnes_2007,Yang_2009,Hai_2009,Yamane_2011,Hayashi_2012,Ieda_2013,Hals_2015,Yamane_2019}. Various nonuniform magnetization textures, such as domain walls (DWs) and skyrmions, are typical research targets for investigating STT and SMF in magnetic materials \cite{Soumyanarayanan_2016}. Magnetic skyrmion also gives rise to the topological Hall effect (THE)\cite{Ye_1999,Tatara_2002,Lee_2009,Neubauer_2009,Kanazawa_2011}. Such concepts realized in dynamical magnetization textures are routinely expressed by the emergent electromagnetic fields acting on electrons\cite{Nagaosa_2012,Nagaosa_2013,Tatara_2019}, whereby new device applications have been proposed\cite{Barnes_2006,Ieda_2012,Nagaosa_2019}. 

This section focuses on the recent advances and perspectives on spin-exchange related phenomena, including spin pumping, topological Hall effect, and emergent inductance.
\subsection{Spin pumping} \label{sec:spin-pumping}
%
%
%
\begin{figure*}[htb]
\begin{center}
\includegraphics[width=16cm]{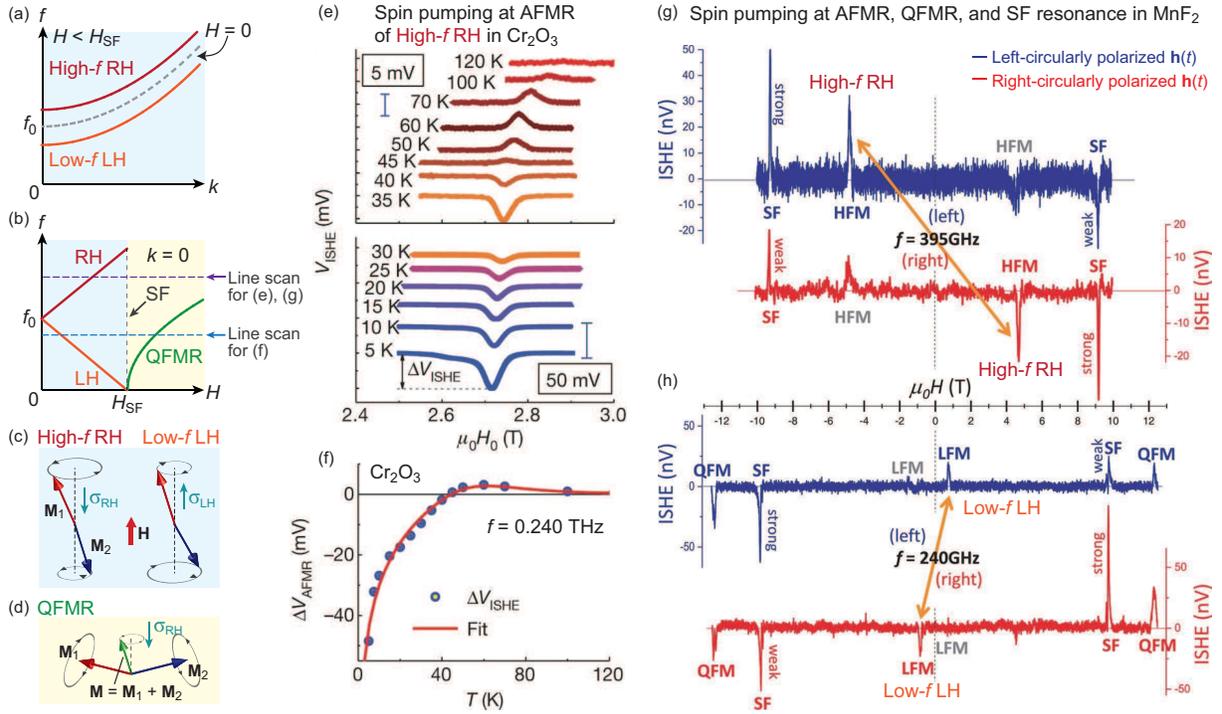}
\end{center}
\caption{
(a) Illustrations of AF spin-wave dispersion relations for an easy-axis antiferromagnet below the spin-flop field, $H_{\rm SF}$. There are two AF branches: high-$f$ right-handed (RH) and low-$f$ left-handed (LH) spin-wave modes with opposite spin polarizations, ${\bm \sigma}_{\rm RH}$ and ${\bm \sigma}_{\rm LH}$, respectively. The modes are degenerate at $f_0$ for $H = 0$ ($f_0 \sim 0.16~\textrm{THz}$ for Cr$_2$O$_3$ \cite{Li2020Nature} and $0.25~\textrm{THz}$ for MnF$_2$ \cite{Vaidya2020Science}). 
(b) Resonance frequency versus $H$ for $k = 0$ ($k$: wavenumber). For $H > H_{\rm SF}$, a quasi-ferromagnetic resonance (QFMR) mode appears. 
(c),(d) Illustrations of (c) high-$f$ RH and low-$f$ LF spin-wave modes for $H < H_{\rm SF}$ and (d) QFMR mode for $H > H_{\rm SF}$. 
(e),(f) ISHE voltage signals at the AFMR of high-$f$ RH mode versus (e) external $H$ and (f) temperature, $T$, when a linearly-polarized microwave at $f =0.240~\textrm{THz}$ is applied to a Cr$_2$O$_3$/(Pt-Ta) sample \cite{Li2020Nature}. 
(g),(h) ISHE voltage signals versus $H$ when a circularly-polarized microwave at (g) $f =0.395$ and (h) $0.240~\textrm{THz}$ is applied to MnF$_2$/Pt samples, which excites the (g) high-$f$ RH and SF and (h) low-$f$ LH, SF, and QFMR modes in MnF$_2$, respectively \cite{Vaidya2020Science}. 
(e),(f) Reproduced with permission from Li {\it et al.}, Nature {\bf 578}, 70--74 (2020). Copyright 2020 Springer Nature Limited. (g),(h) Reproduced with permission from Vaidya {\it et al.}, Science {\bf 368}, 160--165 (2020). Copyright 2020 American Association for the Advancement of Science (AAAS).
}
\label{fig:AF-SP}
\end{figure*}
Spin pumping refers to the generation of a spin current, ${\bf J}_{\rm s}$, from coherent magnetization, ${\bf M}$, precession in a magnet with metallic contact \cite{Mizukami2002PRB,Tserkovnyak2002PRL,Ando2011JAP,Spin-current-text} [Fig. \ref{fig:SP}(a)]. Here, the ${\bf M}$ precession is driven by an applied microwave field, ${\bf h}(t)$, which satisfies the ferromagnetic resonance (FMR) condition. Under FMR, the amplitude of the precessional magnetization motion is resonantly enhanced, and a part of its angular momenta is pumped out of the magnetic layer into a metal attached to the magnet due to interfacial s-d exchange coupling. The generated (conduction-electron) spin current can be detected as a transverse electric voltage in the metal via the inverse spin Hall effect (ISHE) \cite{Saitoh2006APL,Azevedo2005JAP,Valenzuela2006Nature,Costache2006PRL,Kimura2007PRL} when the metal exhibits strong SOC [Fig. \ref{fig:SP}(a)]. Detailed theoretical aspects and basic experimental results of spin pumping are reviewed in Chapter 8 of Ref. \onlinecite{Spin-current-text} and in Refs. \onlinecite{spincurrent_JPSJ} and \onlinecite{Ando2011JAP}. \par
Spin pumping has been extensively studied for a metallic ferromagnet, Ni$_{81}$Fe$_{19}$, \cite{Ando2011JAP,Azevedo2005JAP,Saitoh2006APL,Costache2006PRL} and for a magnetic insulator, Y$_3$Fe$_5$O$_{12}$ (YIG) \cite{Kajiwara2010Nature,Chumak2015NatPhys,Althammer2018JPhysD,Brataas2020PhysRep}. These magnetic materials exhibit high Curie temperatures, low Gilbert damping, and small magnetic anisotropies, which makes their FMR frequencies in the $\sim$ GHz range, accessible easily by experiments. In the past several years, various aspects of spin pumping have been reported using these materials. Some examples include spin pumping driven by parametric excitation \cite{Sandweg2011PRL}, a.c. spin pumping \cite{Hahn2013PRL,Wei2014NatCommun,Hyde2014PRB,Weiler2014PRL}, and spin pumping induced by magnon polaritons \cite{Bai2015PRL,Maier-Flaig2016PRB,Bai2017PRL}, magnon polarons \cite{Hayashi2018PRL}, and magnon-magnon hybrid systems \cite{Li2020PRL_mag-mag,Fan2021AdvMater}.  \par
In the following subsections, we discuss two recent topics related to spin pumping: antiferromagnetic spin pumping and magnetic parametron based on a.c. spin pumping. \par
%
\subsubsection{Antiferromagnetic spin pumping} \label{sec:AFSP}
%
In the emergent field of antiferromagnetic spintronics, spin pumping may offer promising opportunities to access ultrafast magnetization dynamics \cite{Baltz2018RevModPhys,Gomonay2018NatPhys}.
In contrast to ferromagnets, where the excitation frequency, $f$, of ${\bf M}$ dynamics is governed by external ($H$) and anisotropy ($H_{\rm A}$)  fields (typically, $\sim$ GHz frequencies), in antiferromagnetic (AF) materials, the dynamics depends on the combined effects of magnetic anisotropy ($H_{\rm A}$) and the inter-sublattice exchange  ($H_{\rm E}$)  fields, which leads to their excitation frequencies, $f\sim(\gamma/2\pi)\sqrt{2H_{\rm E}H_{\rm A}}$, in the much higher (sub)terahertz range \cite{Baltz2018RevModPhys,Gomonay2018NatPhys,Li2020Nature,Vaidya2020Science}. \par  
In 2020, Li {\it et al.} \cite{Li2020Nature} and Vaidya {\it et al.} \cite{Vaidya2020Science} demonstrated spin pumping from easy-axis antiferromagnets, Cr$_2$O$_3$ and MnF$_2$, having zero-field antiferromagnetic resonance (AFMR) frequency of $f_0 \sim$ 0.16 and 0.25 THz, respectively [depicted schematically in Fig. \ref{fig:SP}(b)].  
In these antiferromagnets below the spin-flop field, $H_{\rm SF}$, there are two different AF spin-wave branches: high-$f$ right-handed (RH) and low-$f$ left-handed (LH) spin precessions with opposite polarizations  \cite{Li2020Nature} [Figs. \ref{fig:AF-SP}(a)-\ref{fig:AF-SP}(d)].  
Li {\it et al.} \cite{Li2020Nature} performed AF spin pumping in Cr$_2$O$_3$/(Pt or Ta) systems with a linearly polarized microwave at 0.240 THz and observed ISHE signals at the AFMR condition for the RH mode [Fig. \ref{fig:AF-SP}(e)]. At a high temperature ($T \sim 100~\textrm{K}$), the polarity of the signal is consistent with that predicted from the spin polarization of the RH mode (${\bm \sigma}_{\rm RH}$), which is the same as that due to the quasi-ferromagnetic (QFM) spin-wave mode \cite{Li2020Nature}  [Figs. \ref{fig:AF-SP}(c) and \ref{fig:AF-SP}(d)]. Interestingly, with decreasing $T$, the sign reverses at around 45 K [Figs. \ref{fig:AF-SP}(e) and \ref{fig:AF-SP}(f)], which may be attributed to incoherent spin pumping (\emph{i.e.}, the spin Seebeck effect) from the LH mode (${\bm \sigma}_{\rm LH}$) due to heating. 
Vaidya {\it et al.} \cite{Vaidya2020Science} conducted AF spin pumping in MnF$_2$/Pt samples with right- and left-circularly polarized microwaves that couple with high-$f$ RH and low-$f$ LH spin-wave modes in MnF$_2$, respectively, in positive fields. 
They found that, when the right- (left-)circularly polarized microwaves are irradiated to the samples, an ISHE voltage with negative (positive) sign appears at 0.395 THz and 4.7 T (0.240 THz and 0.8 T) that satisfies the AFMR condition for the high-$f$ RH (low-$f$ LH) mode in MnF$_2$ [Figs. \ref{fig:AF-SP}(g) and \ref{fig:AF-SP}(h)]. This result agrees with the coherent spin-pumping scenario. They also observed spin-pumping signals at the resonance conditions for SF and QFM spin-wave modes  [Figs. \ref{fig:AF-SP}(g) and \ref{fig:AF-SP}(h)], the characteristic of which is not yet fully understood and requires further investigation from both coherent and incoherent spin dynamics perspectives \cite{Li2020Nature,Vaidya2020Science}. \par 
Spin pumping from different types of antiferromagnets has also been reported recently. 
In 2020, Moriyama {\it et al.} \cite{Moriyama2020PRB} reported AF spin pumping from NiO. Here, NiO is an easy-plane antiferromagnet with the N\'eel temperature ($T_{\rm N}$) of 523 K and shows two linearly polarized magnon modes governed by two different anisotropies \cite{Rezende2019JAP}. 
They measured continuous-wave (cw)-THz wave transmission spectra through NiO-(Pt or Pd) granular systems at $\sim 1~\textrm{THz}$, corresponding to the high-$f$ AFMR mode in NiO arising from the hard axis anisotropy. 
They found that the spectra linewidths for NiO-(Pt or Pd) are broader than that without Pt or Pd, indicating enhanced AF damping caused by spin pumping \cite{Moriyama2020PRB}. 
Soon after, Qiu {\it et al.} \cite{Qiu2021NatPhys} reported THz emission spectroscopy for NiO films covered by Pt or W films and showed ultrafast spin-current generation at zero external field and at room temperature, where a second-order nonlinear optical process in the NiO layer may play an important role. Very recent experimental work by Rongione {\it et al.} \cite{Rongione2022arXiv} revealed that a different (resonant or narrowband) magnon excitation mechanism may involve in THz emission spectra through NiO crystal orientation dependence measurements.
In 2021, Boventer  {\it et al.} \cite{Boventer2021PRL} and Wang  {\it et al.} \cite{Wang2021PRL} independently reported spin pumping from a canted easy-plane antiferromagnet, $\alpha$-Fe$_2$O$_3$, with metallic contacts (Pt or W) as depicted in Fig. \ref{fig:SP}(c). 
Above its Morin transition temperature ($T_{\rm M} \sim 263 ~\textrm{K}$), the Fe$^{3+}$ spins in $\alpha$-Fe$_2$O$_3$ lie in the basal (easy) plane of the corundum lattice and stack antiferromagnetically along the $c$ axis \cite{Wang2021PRL}.  
Here, $\alpha$-Fe$_2$O$_3$ exhibits the bulk Dzyaloshinskii$-$Moriya interaction, which cants the two AF sublattice moments slightly, giving rise to an RH elliptical precessional mode of the net moment in the range of 10 GHz  [Fig. \ref{fig:AF-SP}(d)]. This is much lower than its high-$f$ mode associated with out-of-plane oscillation of the N\'eel vector (0.17 THz). 
Boventer {\it et al.} \cite{Boventer2021PRL} and Wang  {\it et al.} \cite{Wang2021PRL} developed the general framework for spin pumping from this low-$f$ mode and experimentally demonstrated the concept through ISHE measurements. \par 
%
\subsubsection{Towards stochastic computing based on a.c. spin pumping} \label{sec:parametron} 
%
\begin{figure}[tb]
\begin{center}
\includegraphics[width=8.2cm]{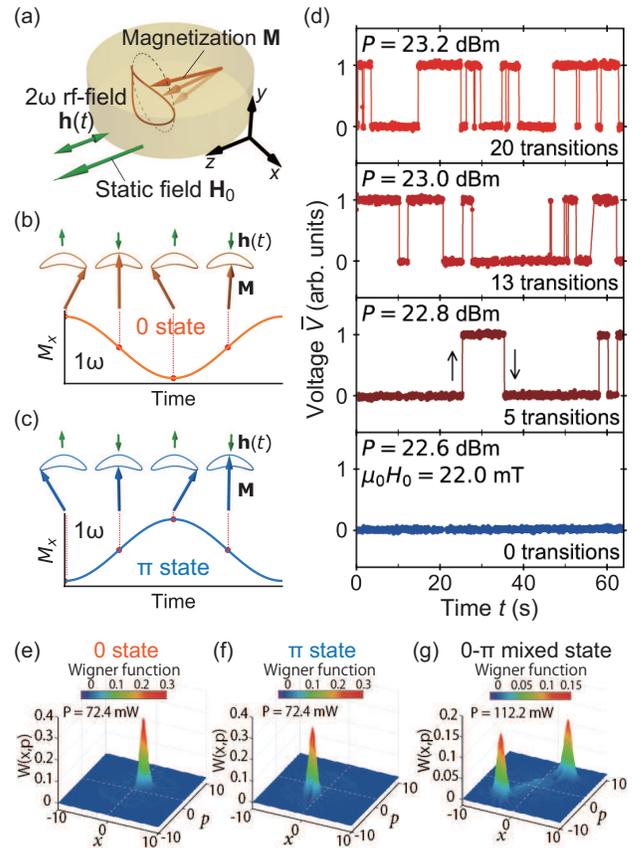}
\caption{(a) Illustration of the parametrically excited magnetization, ${\bf M}$, precession in a YIG disk \cite{Makiuchi2021APL}. (b),(c) Time evolution of $M_x$ (with angular frequency $\omega$) for the (b) 0-phase and (c) $\pi$-phase state that is a relative phase between $M_x$ and the input $2\omega$ microwave field, ${\bf h}(t)$, respectively  \cite{Makiuchi2021APL}. (d) The normalized ISHE voltage, ${\overline V}$, induced by a.c. spin pumping versus time at selected values of microwave power $P$ for a YIG/Pt disk \cite{Makiuchi2021APL}. (e),(f),(g) Reconstructed Wigner function for the (e) 0-phase, (f) $\pi$-phase, and (g) 0-$\pi$ phase mixed state through magnetization state tomography \cite{Hioki2021PRB}. (a)-(d) Reproduced with permission from Makiuchi {\it et al}., Appl. Phys. Lett. {\bf 118}, 022402 (2021). Copyright 2021 AIP Publishing LLC. (e)-(g) Reproduced with permission from Hioki {\it et al}., Phys. Rev. B {\bf 104}, L100419 (2021). Copyright 2021 American Physical Society.} \label{fig:Magnetic_parametron}
\end{center}
\end{figure}
Electric readout of spin information based on spin pumping and ISHE may provide a potential pathway for future computational technologies. 
In 2021, Makiuchi {\it et al.} \cite{Makiuchi2021APL} built a parametric oscillator based on a YIG disk that acted as a ``magnetic parametron'', a nonlinear magnetic oscillator whose precessional  phases are discretized into Ising-like $0$ or $\pi$ relative to the applied microwave phase (Fig. \ref{fig:Magnetic_parametron}). In their experiment \cite{Makiuchi2021APL}, the binary phase of magnetization, ${\bf M}$, precession (with angular frequency $\omega$) was realized by parallel parametric pumping in the YIG disk with shape magnetic anisotropy; when a $2\omega$ microwave for the parametric excitation was applied, one of the energetically-degenerated phase states spontaneously selected  [Figs. \ref{fig:Magnetic_parametron}(b) and \ref{fig:Magnetic_parametron}(c)]. The phase information can be electrically read out through a.c. spin pumping and ISHE measurements [see Fig. \ref{fig:SP}(a)] in a Pt film attached to the YIG disk at room temperature. Makiuchi {\it et al.} \cite{Makiuchi2021APL} showed that the phase states undergo transition between stable and stochastic regimes by increasing the excitation power [Fig. \ref{fig:Magnetic_parametron}(d)]. In the latter regime, they further demonstrated that the occurrence probability of each state can be tuned with additional microwaves, showing its potential application as a ``probabilistic bit (p-bit)'' in stochastic computation.  
By combining the above experimental technique with that established in quantum optics, Hioki {\it et al.} \cite{Hioki2021PRB} demonstrated the state tomography for magnetization dynamics. They obtained a density matrix and Wigner function realized in a magnetic parametron, which shows a mixed state composed of two coherent states [Figs. \ref{fig:Magnetic_parametron}(e)-\ref{fig:Magnetic_parametron}(g)]. 
Shimizu {\it et al.} \cite{Shimizu2022APL} numerically studied spin dynamics in a magnetic parametron based on a master equation and found that amplitude squeezed states can be formed under strongly biased microwaves. 
More recently, Elyasi {\it et al.} \cite{Elyasi2021arXiv} theoretically showed that there are three dynamical phases in a magnetic parametron: a stable Ising spin, telegraph noise of thermally activated switching, and an intermediate regime that at lower temperatures is quantum-correlated with significant distillable magnon entanglement. Their finding further expands the scope of the magnetic parametron application as a quantum information processor \cite{Chumak2022IEEE}. \par 

\subsection{Topological Hall torque}
\begin{figure}[tbh]
\begin{center}
\includegraphics[width=8.5cm]{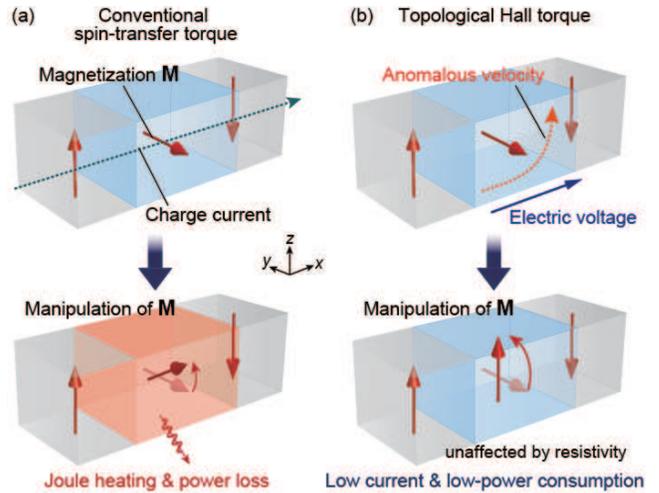}
\caption{
Illustrations of the (a) conventional spin-transfer torque (STT) exerted on a magnetic texture by a dissipative charge current and (b) topological Hall torque (THT) emerging via nondissipative anomalous Hall current by applying an electric voltage. 
}\label{fig:Topological_Hall_torque}
\end{center}
\end{figure}

Electrical manipulation of magnetization is important for establishing next-generation magnetic storage, such as magnetoresistive random access memory (MRAM). To this end, electrically induced torques exerting on magnetic textures has been considered. Two types of such torques were identified: STT\cite{Slonczewski_1989,Berger_1996, Myers_1999} [Fig. \ref{fig:Topological_Hall_torque}(a)] and spin-orbit torque (SOT)\cite{Obata_2008,Manchon_2008,Miron_2010,Miron_2011,Liu_2012,Liu_2012_2}, aiming for a high-speed and energy-efficient writing scheme for MRAM. The effects are proportional to the charge-to-spin conversion ratio, and thus highly spin-charge coupled systems are pursued for energy-efficient DW motion. However, they face a physical limit of the charge-to-spin conversion ratio, which is a fundamental obstacle in those conventional mechanisms. To circumvent this problem, the use of topological physics, particularly, Weyl electrons emerging around Weyl points (WPs), can be employed because electrons acquire a highly efficient charge-to-spin conversion thanks to a large fictitious magnetic field (Berry curvature) and spin-momentum locking (SML).

Momentum-space Berry curvature leads to an anomalous velocity transverse to the applied electric field, which is the origin of the intrinsic anomalous Hall effect (AHE)\cite{Karplus_1954,Sundaram_1999,Nagaosa_2006,Sinitsyn_2007,Nagaosa_2010}.
In the vicinity of the band inversion point associated with a strong SOC 
on the surface of topological insulators (TIs) \cite{Nomura_2011,Yu_2010,Chang_2013,Checkelsky_2012} and in Weyl semimetals (WSMs) \cite{Grushin_2012,Goswami_2013,Burkov_2014,Burkov_2014_2,Nakatsuji_2015,Nakatsuji_2016,Nayak_2016,Liu_2018,Wang_2018},
the Berry curvature becomes significant and yields a large anomalous Hall conductivity.
Relying on the strong SOC, the electrically induced torques were studied within particular models of TIs \cite{Garate_2010,Yokoyama_2010,Pesin_2012,Tserkovnyak_2012,Sakai_2014,Ndiaye_2017,Kurebayashi_2019,Imai_2021} and WSMs \cite{Kurebayashi_2016,Kurebayashi_2017,Kurebayashi_2019_2}. In a recent study, a classification of the contributions to the torques based on the semiclassical (Boltzmann) formalism has been established\cite{Araki_2021}, revealing the emergence of intrinsic torques driven by the anomalous velocity via SML. 
In comparison with the conventional STT\cite{Slonczewski_1989,Berger_1996,Myers_1999} [Fig. \ref{fig:Topological_Hall_torque}(a)] and SOT\cite{Obata_2008,Manchon_2008,Miron_2010,Miron_2011,Liu_2012,Liu_2012_2} induced by transport current, the intrinsic torques are robust against disorder or thermal fluctuation.  The inverse spin-galvanic effect was proposed for the surface states of topological insulators attached to ferromagnets\cite{Garate_2010} and has been demonstrated experimentally\cite{Wu_2021}. Even in bulk crystals without breaking of inversion symmetry by surfaces or interfaces, an intrinsic torque is generated by the coupling of momentum-space Berry curvature with real-space magnetic textures\cite{Araki_2021}, which is called the ``topological Hall torque (THT)'' [Fig. \ref{fig:Topological_Hall_torque}(b)].  

The structure of the THT is formally compatible with the field-like STT, albeit it does not rely on the transport spin current flowing through magnetic textures, such as DWs. Recently, an interaction between a DW and Weyl electrons has been investigated in a ferromagnetic oxide, SrRuO$_3$ \cite{Yamanouchi2022}, which has many WPs near the Fermi level. Although the current density required for the DW motion in SrRuO$_3$ is more than one order of magnitude lower than that in metallic systems\cite{Feigenson2007,Yamanouchi2019}, its mechanisms have been unsolved until Ref.~\onlinecite{Yamanouchi2022}. When a current is applied across the DW, an effective magnetic field, $H_\mathrm{eff}$, is exerted on the DW. The ratio of $H_\mathrm{eff}$  per current density is ``over one order of magnitude higher (more efficient)'' than that originating from conventional STT and SOT reported to date. Within a framework of THT, by inserting the values typical in SrRuO$_3$, the magnitude of $\mu_0H_\mathrm{eff}/J$ is $10-12$ Tm$^2/$A, which formally corresponds to the nonadiabaticity parameter, $\beta_\mathrm{THT}\simeq 2$, and matches the measured value.

Although effects of Weyl electrons appear in a wide variety of phenomena, such as magnetotransport\cite{Grushin_2012,Goswami_2013,Burkov_2014,Burkov_2014_2,Nakatsuji_2015,Nakatsuji_2016,Nayak_2016,Liu_2018,Wang_2018}, spin wave\cite{Itoh_2016}, quantum phenomena, and thermoelectric phenomena, the findings in Ref.~\onlinecite{Yamanouchi2022} add DW dynamics to that catalog and pave the way for an energy-efficient scheme for electrical manipulation of magnetization, which is essential in the operation of next-generation magnetic storage devices. Prominently, the mechanism reported in Ref.~\onlinecite{Araki_2021} is not limited to the special class of Weyl semimetals, but appears in a ferromagnetic metal with WPs, which is a more general material. The new torque mechanism, THT, therefore, offers applications of topological physics, one of the major topics in current solid-state physics, to spintronics. 

\subsection{Emergent inductor}
\begin{figure}[tb]
\begin{center}
\includegraphics[width=8.5cm]{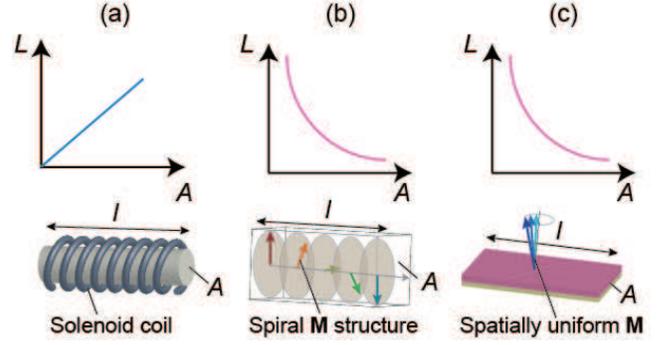}
\caption{
Schematics of the inductance, $L$, versus cross-sectional area, $A$, (upper panels) for the (a) conventional inductor, (b) emergent inductor based on a spiral magnet, and (c) spin-orbit emergent inductor based on a collinear magnet showing SOC (lower panels). $l$ and $M$ denote the length and magnetization, respectively.
} \label{fig:Inductors}
\end{center}
\end{figure}

Inductors are basic components in electric circuits realizing functions such as voltage transformation, noise filtering, switching, etc.
The mechanism relies on classical electromagnetics: a conducting coil stores the energy in a magnetic field when an electric current flows through it, leading to the induction of an electromotive force that opposes the current change.
An inductance of a solenoid coil [Fig. \ref{fig:Inductors}(a)] is given by $ L=\mu n^2 l A $, where $\mu,\,n,\,l,\,$ and $A$ are the coil permeability, turn density, length, and cross section, respectively. This relation indicates that the magnitude of the inductance scales with the size of the solenoid, which limits downsizing inductors in microcircuits. 

Because the SMF is the spin version of Faraday's law of induction, one can envisage the spin-extension of inductor operation in magnetic nanostructures.
This was actually invented by a theoretical proposal of the so-called ``emergent inductor'' using a spiral magnet\cite{Nagaosa_2019} [Fig. \ref{fig:Inductors}(b)].
When an electric current flows in a spiral magnet, it stores the energy in the spiral structure formed by the local magnetization via its exchange coupling with conduction electrons.
In terms of the Berry phase formalism, where the spin Berry phase is given by the solid angle sustained by magnetization dynamics\cite{Stern_1992, Nagaosa_2012, Nagaosa_2013,Barnes_2007}, the emergent inductance can be thought of as an extension of the dynamical Aharonov-Bohm effect to a spiral magnet, where the electromagnetic potential is replaced by a spin-dependent Berry connection generated by the spatial variation of the magnetization\cite{Berger_1986,Volovik_1987,Barnes_2007}.
Focusing on the adiabatic processes of STT and SMF, the inductance originating from spiral dynamics can be expressed as the following simple formula,
\begin{equation}
  L =  \left(\frac{ p \hbar  q}{ 2 e  \sqrt{K}}\right)^2 \frac{ l}{ A }, 
\label{l}
\end{equation} 
where $p$ is the spin polarization, $\hbar$ is the Dirac constant, $q$ is the spiral wave number, $e$ is the elementary charge, $K$ is the magnetic hard axis anisotropy constant, $l$ is the spiral magnet length, and $A$ is the spiral magnet cross-sectional area.
A prominent feature of emergent inductors is their inductance magnitude dependence on physical size.
Contrary to the coil inductance that is proportional to the coil cross section area, $A$ ( $L\propto A$), the emergent inductance is inversely proportional to the area that the current passes through ($L\propto A^{-1}$).
This property opens an innovative avenue for downsizing inductor elements.

Soon after the theoretical proposal, the concept was experimentally demonstrated in a centrosymmetric helical magnet, Gd$_3$Ru$_4$Al$_{12}$\cite{Yokouchi}.
More recent theoretical studies have shown that two excitation modes of a spiral magnetic texture, namely its translational displacement and rotation of the spiral plane, contribute to emergent inductance with opposite signs\cite{Ieda_2021,Kurebayashi}.
This may explain the negative inductance observed in the experiment\cite{Yokouchi}, whereas the original theory, which takes only into account the rotational excitation, predicted positive inductance\cite{Nagaosa_2019}.
Experimentally, a room temperature observation of emergent inductance has been achieved in YMn$_6$Sn$_6$\cite{Kitaori}, whereas in Ref.~\onlinecite{Yokouchi} the temperature had to be as low as below $\sim20$ K using Gd$_3$Ru$_4$Al$_{12}$.
The discovery of emergent inductance has reopened electronics theories and we are at the beginning of a new chapter exploring inductance by quantum mechanical mechanisms.

The concept of the emergent inductor is not limited to original spiral dynamics. In fact, a novel inductance of SOC origin was proposed\cite{Yamane_2022} where an SOC stores the energy in itself and mediates the energy conversion with the electric energy.
The spin-orbit inductance results from the time derivative of the Aharonov-Casher phase\cite{Aharonov} in magnetic materials [Fig. \ref{fig:Inductors}(c)], where the Berry connection originating from SOCs depends on the electron's momentum and spin.
As already shown in Ref.~\onlinecite{Ieda_2021}, effects of a Rashba-type SOC play significant roles in a spiral-based emergent inductor.
The spin-orbit inductance can be formulated based on a dynamical spin Berry phase acquired by an electron moving in arbitrary magnetic textures in the presence of SOCs in a general form.
\begin{equation}
	L_{ i j } = \frac{ p m_{\rm e} }{ e } \frac{ l_i }{ A } \sum_k g_{ i k }  \chi_\omega^{ k j } ,
	\label{lso}
\end{equation}
where $m_\mathrm{e}$ is the electron mass, $ l_i $ is the dimension of the sample in the $ x_i $ direction, $ A $ is the cross sectional area normal to the electric current, $g_{ i j } $ is the general SOC, and $ \chi_\omega^{ i j }$ is the magnetic susceptibility tensor with respect to current-induced torques\cite{}.
Exploring spin-orbit inductance with spatially uniform magnetization is of particular interest, where the other inductance mechanisms are ruled out.
With the ferromagnetic resonance frequency, $ \omega_{\rm R} $, defined by $ \omega_{\rm R} \propto K$ ($K$ is the hard axis anisotropy constant), the dynamical susceptibility for $ \omega \ll \omega_{\rm R} $ can be approximated by $ (  \chi_\omega^{ x x } ,  \chi_\omega^{ y x } ) \simeq - \frac{ p m_{\rm e} }{ 2 e K } ( g_{ x x} , g_{ x y } ) $.
In this low frequency regime, the inductances in Eq.(\ref{lso}) can be approximated by
\begin{eqnarray}
	L_{ x x } &=& \left( \frac{ p m_{\rm e} }{ \sqrt{ 2 } e } \right)^2 \frac{ l_x }{ A }
	                     \frac{ g_{ x x}^2 + g_{ x y }^2 }{ K } , \label{l0} \\
	L_{ y x } &=& \left( \frac{ p m_{\rm e} }{ \sqrt{ 2 } e } \right)^2 \frac{ l_y }{ A }
	                       \frac{ g_{ y x } g_{ x x } + g_{ y y } g_{ x y } }{ K } . \label{ly}
\end{eqnarray}
Note that spin-orbit inductance appears in both the longitudinal and transverse (Hall) directions with respect to a current, and they are the second order of the SOC constants $ g_{ i j } $.

From a technological perspective, we compare classical, spiral-based, and spin-orbit inductances as follows. The classical inductance scales with the size of the coil as schematically shown in Fig. \ref{fig:Inductors}(a), making it highly challenging to miniaturize an inductor.
The spiral-based emergent inductance, originating from the quantum-mechanical exchange coupling effect, is free from the undesirable system-size dependence. It is inversely proportional to the system's cross sectional area (normal to the electric current direction)\cite{Nagaosa_2019}, as illustrated in Fig. \ref{fig:Inductors}(b). 
In Ref.~\onlinecite{Yokouchi}, an emergent inductance has been reported comparable in its magnitude to that of a commercial one ($ \sim 400 $ nH), but in a volume about a million times smaller.
Although the emergent inductance has broken a hurdle for manufacturing smaller inductors with larger effects, it has a limitation in the operating frequency.
The previous experiments successfully observed the emergent inductance only up to the frequencies of sub-megahertz\cite{Kitaori} and megahertz\cite{Yokouchi} because a spiral magnetic structure cannot respond collectively and robustly to electric currents with higher frequencies.
Spin-orbit inductance resolves the two issues, regarding miniatuarization and the operating frequency, simultaneously.
As seen in Eq.~(\ref{lso}), the smaller the cross sectional area, $ A $, the larger the spin-orbit inductance, as is the case for the spiral-based emergent inductance.
Spin-orbit inductance with spatially uniform magnetization also provides nearly frequency independent real parts, except in the vicinity of the resonance frequency, which is typically $\sim$ 1--10 GHz. 
Potential candidate systems for experimental observation of spin-orbit inductance include heavy metal/ferromagnet heterostructures, as shown in Fig. \ref{fig:Inductors}(c), where a Rashba SOC arises due to the structural inversion asymmetry.
Those systems have been extensively studied in spintronics for use in non-volatile memory devices\cite{Miron_2010,Miron_2011,Liu_2012,Liu_2012_2}.
Adopting $ g = 10^{-10} $ eV$\cdot$m/$\hbar$ \cite{Miron_2010} and employing some typical values for the other relevant material parameters as $ K \sim 10^5$ J/m$^3$, $ p \sim 0.5 $, and the bare electron mass for $ m_{\rm e} $, $ L_{ x x } $ in Eq.~(\ref{l0}) is estimated as $ \sim 10^{-18} \times ( l_x / A ) $ H.
Assuming the sample dimensions of $ ( l_x, l_y, l_z ) = ( 0.1{\rm mm}, 100{\rm nm}, 10{\rm nm} ) $, we calculate $ L_{ x x } \sim 100 $ nH.
In the context of conventional spintronics applications, larger $ K $ has been mostly pursued for better thermal stability of the magnetic configuration\cite{Dieny_2020}. 
For a larger spin-orbit inductance, in contrast, smaller $ K $ is preferred, see Eqs.~(\ref{l})--(\ref{ly}), while it also leads to a lower resonance frequency.
It is desired to systematically conduct experimental and material research to achieve optimal conditions for spin-orbit inductance, \emph{i.e.}, simultaneous realization of larger $ g_{ i j } $, smaller $ A $, and the right magnitude of $ K $.
 
%
%
\section{\label{Spin-heat coupling}Spin-heat coupling}
We now extend our discussion to the interaction/interconversion of spins with heat, lattice vibrations, and charge current. Starting with a brief introduction and study summaries of spin-lattice coupling, we address recent progress and perspectives on the spin Seebeck and Peltier effects, two representative spin-heat coupling phenomena. \par 
%
\subsection{Spin-lattice coupling} \label{sec:spin-lattice}
\begin{figure*}[bht]
\begin{center}
\includegraphics[width=12cm]{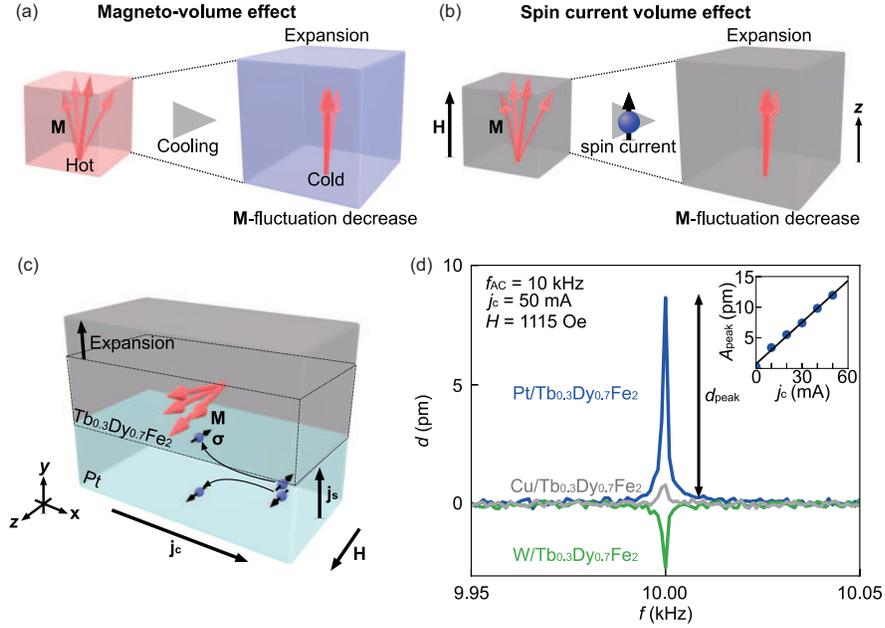}
\end{center}
\caption{
(a) Illustration of the magneto-volume effect (MVE). A ferromagnet expands (shrinks) via the spin-lattice coupling when spin fluctuation in the magnet decreases (increases) due to the magnetic-field, ${\bf H}$, application or temperature modulation \cite{Arisawa2022NatCommun}. 
(b) An illustration of the spin-current volume effect (SVE), where the volume of a ferromagnet is tuned by spin-current injection \cite{Arisawa2022NatCommun}. 
(c) A schematic of the SVE induced by the SHE in a Pt/Tb$_{0.3}$Dy$_{0.7}$Fe$_2$ system \cite{Arisawa2022NatCommun}. 
When a charge current, ${\bf j}_{\rm c}$, is applied to the Pt film, a spin current, ${\bf j}_{\rm s}~||~{\bf {\hat y}}$ (with spin polarization ${\bm \sigma}~||~{\bf {\hat z}}$), is injected into the Tb$_{0.3}$Dy$_{0.7}$Fe$_2$ film, and magnetization, ${\bf M}$, fluctuation in the Tb$_{0.3}$Dy$_{0.7}$Fe$_2$ film decreases. This causes volume expansion via spin-lattice coupling, which accompanies a thickness change of the Tb$_{0.3}$Dy$_{0.7}$Fe$_2$ film \cite{Arisawa2022NatCommun}.  
(d) Mechanical vibrational spectrum (the signed amplitude of the vibration $d$) for (Pt, W, Cu)/Tb$_{0.3}$Dy$_{0.7}$Fe$_2$ samples measured with laser Doppler vibrometry. 
The inset shows the $j_{\rm c}$ dependence of the mechanical vibration amplitude, $A_{\rm peak}$, at $H=1530~\textrm{Oe}$ \cite{Arisawa2022NatCommun}. 
Reproduced from Arisawa {\it et al.}, Nat. Commun. {\bf 13}, 2440 (2022). Copyright 2022 Author(s), licensed under a Creative Commons Attribution (CC BY) license.
} \label{fig:SVE}
\end{figure*}

%
Spin-lattice coupling is a fundamental mechanism of spin-heat coupling and responsible for various phenomena that change both static and dynamical magnetization/lattice states, such as magnetostriction, magneto-volume effect (MVE) [Fig. \ref{fig:SVE}(a)], and magnon-phonon interaction (leading to magnon-phonon thermalization). 
The mechanism can be classified into two-ion and single-ion types \cite{Gurevich-Melkov_text,Keffer-text,Luthi_text}.  
The former arises due to the change of the exchange interaction ($J$) between magnetic atoms (or ions) when their distances are modulated by lattice vibrations, whereas the latter is due to the strain-induced change of the magnetic anisotropy, or spin-orbit interaction \cite{Gurevich-Melkov_text,Keffer-text,Luthi_text}. \par
Recently, spin-lattice coupling has renewed interest in spintronics.
By the single-ion type interaction, magnons and phonons in the vicinity of the crossings of their dispersion relations, are hybridized into quasiparticles called ``magnon polarons'' that share mixed magnonic and phononic characters \cite{Kamra2015PRB,Shen2015PRL,Kikkawa2016PRL,Flebus2017PRB}. Magnon polarons can convey spin information with velocities close to those of phonons, much faster than the magnon velocities in the dipolar regime \cite{Shen2015PRL,Frey2021PRB}. 
Owing to the long-lived phononic constituent, magnon polarons may have longer lifetimes than pure magnons and can enhance the spin-current related phenomena, such as spin pumping \cite{Hayashi2018PRL} (Sec. \ref{sec:spin-pumping}) and spin Seebeck/Peltier effects \cite{Kikkawa2016PRL,Flebus2017PRB,Yahiro2020PRB} (Secs. \ref{sec:SSE} and \ref{sec:SPE}). 
Magnon-phonon interconversion has also been detected through Brillouin light scattering \cite{Holanda2018NatPhys}, FMR \cite{An2020PRB}, and spatio-temporal imaging \cite{Hioki2020ComPhys} spectroscopy, which may open a new avenue for empowering information transfer with magnon polarons in quantum transduction devices \cite{Li2021APLMater,Rameshti2022PhysRep}. \par 
To further utilize spin-lattice coupling in spintronics, spin-current injection into a large magnetostrictive material may offer a unique opportunity. In 2022, Arisawa {\it et al.} showed that volume of a magnet can be manipulated by injecting a spin current: the spin-current volume effect (SVE) [Fig. \ref{fig:SVE}(b)] \cite{Arisawa2022NatCommun}. They observed that the thickness of thin films of ferromagnetic Tb$_{0.3}$Dy$_{0.7}$Fe$_2$ exhibiting strong spin-lattice coupling changes by a spin current induced by the spin Hall effect (SHE) of the attached Pt or W films [Figs. \ref{fig:SVE}(c) and \ref{fig:SVE}(d)]. Theoretical calculation revealed that modulation of magnetization fluctuation due to the spin-current injection plays an essential role \cite{Arisawa2022NatCommun}.
The SVE enables direct mechanical actuation of a magnetostrictive thin film by using a spin current, and may provide a promising approach to explore spintronic phenomena driven by spin-lattice coupling. \par  
%
\subsection{Spin Seebeck effect} \label{sec:SSE}
\begin{figure}[tb]
\begin{center}
\includegraphics[width=8.5cm]{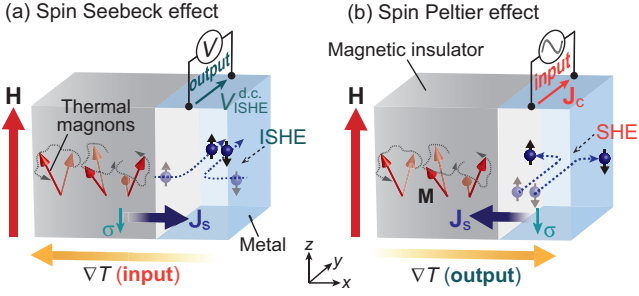}
\caption{(a),(b) Illustrations of the (a) SSE and (b) SPE in a magnetic insulator with metal contact. 
(a) For the SSE, an applied temperature gradient, $\nabla T~(||~{\bf {\hat x}})$, excites a magnon flow, which at the interface to the metal becomes a conduction-electron spin current, ${\bf J}_{\rm s}~||~{\bf {\hat x}}$ (with spin polarization ${\bm \sigma}~||~{\bf H}~||~{\bf {\hat z}}$), and is converted into a transverse voltage by the ISHE according to the relation ${\bf J}_{\rm s} \times {\bm \sigma}$. 
(b) For the SPE, an applied charge current, ${\bf J}_{\rm c}~(||~{\bf {\hat y}})$, generates a conduction-electron spin current, ${\bf J}_{\rm s}~||~{\bf {\hat x}}$ (with spin polarization ${\bm \sigma}~||~{\bf {\hat z}}$), by the SHE, which at the interface to the magnet is converted into a magnon flow. Due to energy conservation, this process includes heat transfer between the electron and magnon systems, leading to a measurable temperature difference between the metal and magnet. 
}\label{fig:SSE-SPE}
\end{center}
\end{figure}

The spin Seebeck effect (SSE) \cite{Uchida2014JPCM,Uchida2016ProcIEEE,Kikkawa2023ARCMP} refers to the generation of a spin current, ${\bf J}_{\rm s}$, as a result of a temperature gradient, $\nabla T$, in magnetic materials.
It is well established for magnetic insulators with metallic contacts, at which a magnon spin current is converted into a conduction-electron spin current by the interfacial s-d exchange coupling and detected as an ISHE voltage \cite{Xiao2010PRB,Adachi2013RepProgPhys,Rezende2014PRB,Barker2016PRL,Rezende_text}  [Fig. \ref{fig:SSE-SPE}(a)].  
In particular, the YIG/Pt heterostructure \cite{Uchida2010APL} has become a prototype system.   
Ferrimagnetic YIG exhibits the smallest magnetic damping, high Curie temperature ($T_{\rm C} \sim 560~\textrm{K}$), and high resistivity owing to a large band gap \cite{Qiu2015APEX,Althammer2018JPhysD}, whereas Pt is a paramagnetic metal showing high ISHE efficiency \cite{Hoffman2013IEEE,Sinova2015RMP}, which is ideal for SSE measurements.  
A number of experiments have been conducted using YIG and YIG/Pt systems to reveal the physics behind SSEs.
Reports include 
temperature \cite{Uchida2014PRX,Kikkawa2015PRB,Jin2015PRB,Guo2016PRX,Iguchi2017PRB}, 
magnetic field \cite{Uchida2015PRB,Kikkawa2015PRB,Jin2015PRB,Kikkawa2016PRL,Cornelissen2016PRB_H-dep,Guo2016PRX,Kikkawa2016JPSJ,Miura2017PRMater}, 
length-scale (thickness) \cite{Kikkawa2015PRB,Jin2015PRB,Giles2015PRB,Kehlberger2015PRL,Cornelissen2015NatPhys,Guo2016PRX,Miura2017PRMater,Prakash2018PRB}, 
structure \cite{Wu2018PRL,Nozue2018APL},
time \cite{Agrawal2014PRB,Roschewsky2014APL,Hioki2017APEX,Bartell2017PRAppl,Kimling2017PRL,Seifert2018NatCommun} dependence measurements,  
separation with other (thermo)electric and spin-current effects \cite{Kikkawa2013PRL,Qu2013PRL,Kikkawa2013PRB,Schreier2014JPhysD,Vlietstra2014PRB,Miao2016AIPAdv,Kikkawa2017PRB,Chang2017PRMater,Avci2015APL}, 
quantitative estimation of SSE thermoelectric coefficient \cite{Sola2017SciRep,Sola2017IEEE,Venkat2020RevSciInst},
neutron scattering experiments to determine magnon dispersions to unraveling SSE features \cite{Man2017PRB,Princep2017npjQM,Shamoto2018PRB,Nambu2020PRL}, 
evaluation of a magnon temperature and chemical potential \cite{Agrawal2013PRL,An2016PRL,Cornelissen2016PRB_chemical-potential,Rezende_text,Olsson2020PRX}, and so on. 
Detailed experimental and theoretical aspects of SSEs are reviewed in Chapter 18 of Ref. \onlinecite{Spin-current-text} and in Refs. \onlinecite{Uchida2014JPCM}, \onlinecite{Uchida2016ProcIEEE}, \onlinecite{Kikkawa2023ARCMP}, and \onlinecite{Adachi2013RepProgPhys}. \par
In the following subsections, we will give an overview and perspective about the selected recent topics in SSE research: antiferromagnetic SSEs, SSEs in van der Waals two-dimensional (2D) materials, and SSEs driven by magnon polarons, quantum spins, and nuclear spins. \par
%
\subsubsection{Antiferromagnetic spin Seebeck effect} \label{sec:AF-SSE}
The SSE provides a much simpler method to generate a spin current through AF spin dynamics, as compared to coherent spin pumping, which needs microwave instruments compatible with a subterahertz frequency range \cite{Li2020Nature,Vaidya2020Science}. 
Since 2015, AF SSEs have been reported for various materials, including 
Cr$_2$O$_3$ \cite{Seki2015PRL,Yuan2018SciAdv,Li2020Nature,Li2020PRL_Cr2O3,Yuan2020APL,Luo2021PRB,Muduli2021AIPAdv}, 
MnF$_2$ \cite{Wu2016PRL}, 
FeF$_2$ \cite{Li2019PRL}, 
$\alpha$-Fe$_2$O$_3$ \cite{Lebrun2018Nature,Yuan2020APL,Ross2021PRB}, 
NiO \cite{Holanda2017APL,Ribeiro2019PRB,Gray2019PRX,Hoogeboom2020PRB}, 
$\alpha$-Cu$_2$V$_2$O$_7$ \cite{Shiomi2017PRB}, 
MnPS$_3$ \cite{Xing2019PRX_MnPS3,Chen2021NatCommun_MnPS3}, 
SrFeO$_3$ \cite{Hong2019APL}, 
SrMnO$_3$ \cite{Das2021APL}, 
YFeO$_3$ \cite{Das2022NCOM}, 
LaFeO$_3$ \cite{Lin2022NatPhys}, 
LuFeO$_3$ \cite{Xu2022PRL}, 
DyFeO$_3$ \cite{Hoogeboom2021PRB}, and 
electrically-switchable BiFeO$_3$ \cite{Parsonnet2022PRL}, showing the versatility of the SSE to investigate AF spin dynamics.  
There is, however, still debate on the $H$ response for easy-axis antiferromagnets below $H_{\rm SF}$.
Seki {\it et al.} \cite{Seki2015PRL} did not observe any detectable signal in Cr$_2$O$_3$/Pt, whereas Wu {\it et al.} \cite{Wu2016PRL} observed  a ferromagnetic (positive) sign \cite{Schreier2014JPhysD}. Recently, Li {\it et al.} \cite{Li2020Nature,Li2020PRL_Cr2O3,Yuan2020APL} reported a negative sign for Cr$_2$O$_3$/Pt and $\alpha$-Fe$_2$O$_3$/Pt in its easy-axis AF phase ($T<T_{\rm M}$), which is consistent with the spin polarization carried by the low-$f$ LH mode [Figs. \ref{fig:AF-SP}(a)-\ref{fig:AF-SP}(c)].  
The magnon-polaron SSE anomalies observed in Cr$_2$O$_3$/Pt corroborate this scenario \cite{Li2020PRL_Cr2O3}.  
Li {\it et al.} \cite{Li2020Nature,Yuan2020APL} further showed that the SSE sign for $H < H_{\rm SF}$ changes from negative to positive when the surface of Cr$_2$O$_3$ is etched before Pt deposition. 
The result may be interpreted in terms of the appearance of uncompensated magnetic moments at the interface that contribute to the positive ferromagnetic-like SSE signal through modification of the interfacial spin-mixing conductance or generation of an additional spin current \cite{Li2020Nature,Yuan2020APL}. 
The sign of the AF SSE for $H < H_{\rm SF}$ is also a subject of theoretical investigations \cite{Rezende2016PRB,Yamamoto2019PRB,Reitz2020PRB,Yamamoto2022PRB}. 
The interfacial and bulk magnon transport theories for AF SSEs predict the negative sign due to greater thermal occupation of the low-$f$ LH mode \cite{Rezende2016PRB,Reitz2020PRB}. By contrast, the Landau-Ginzburg theory \cite{Yamamoto2019PRB} near $T_{\rm N}$ predicts the same sign as ferromagnets. Further experimental and theoretical studies would therefore be desirable to elucidate the mechanisms behind the AF SSEs. 
Very recently, Yamamoto {\it et al.} \cite{Yamamoto2022PRB} theoretically addressed the sign issue and showed that the negative (positive) sign appears for $H < H_{\rm SF}$ under the condition that the interfacial coupling between the conduction-electron spins, ${\bf s}$, and the N\'eel order, ${\bf n}$ (net magnetization ${\bf m}$), of the AF layer dominates the interfacial spin-current generation. \par
\subsubsection{Spin Seebeck effect in two-dimensional (2D) materials} \label{sec:2D-SSE}
Van der Waals 2D materials may also be an intriguing platform for studying  SSEs. 
Ito {\it et al.} \cite{Ito2019PRB} studied the SSE in quasi-2D layered ferromagnets, Cr$_2$Si$_2$Te$_6$ ($T_{\rm C} \sim 31~\textrm{K}$) and Cr$_2$Ge$_2$Te$_6$ ($T_{\rm C} \sim 65~\textrm{K}$), with Pt contacts. These 2D materials show in-plane short-range ferromagnetic correlations, which survive up to at least 300 K (for Cr$_2$Si$_2$Te$_6$), whereas out-of-plane correlations disappear slightly above $T_{\rm C}$ \cite{Williams2015PRB}. 
The SSE in these systems turned out to persist above $T_{\rm C}$, which may be attributed to exchange-dominated interlayer transport of in-plane paramagnetic moments reinforced by short-range ferromagnetic correlations and strong Zeeman effects \cite{Ito2019PRB}. 
Magnon transport in 2D magnets, such as antiferromagnetic MnPS$_3$ \cite{Xing2019PRX_MnPS3,Chen2021NatCommun_MnPS3} and ferromagnetic CrBr$_3$ \cite{Liu2020PRB_CrBr3}, has also been investigated via nonlocal SSEs.  
In 2020, Lee {\it et al.} \cite{Lee2020AFM_WSe2} showed that, when a monolayer WSe$_2$ is inserted between Pt and YIG layers, the SSE is enhanced by a factor of $\sim 5$ compared to that in a Pt/YIG system, which may offer a new opportunity in SSE research with 2D transition dichalcogenide materials. 
SSE research using 2D materials is still in its infancy, but more intriguing results will emerge with the discovery of new functional 2D materials and stacking combinations \cite{Kikkawa2023ARCMP}. \par
%
\subsubsection{Magnon-polaron spin Seebeck effect} \label{sec:mpSSE}
%
%
\begin{figure}[tb]
\begin{center}
\includegraphics[width=8.2cm]{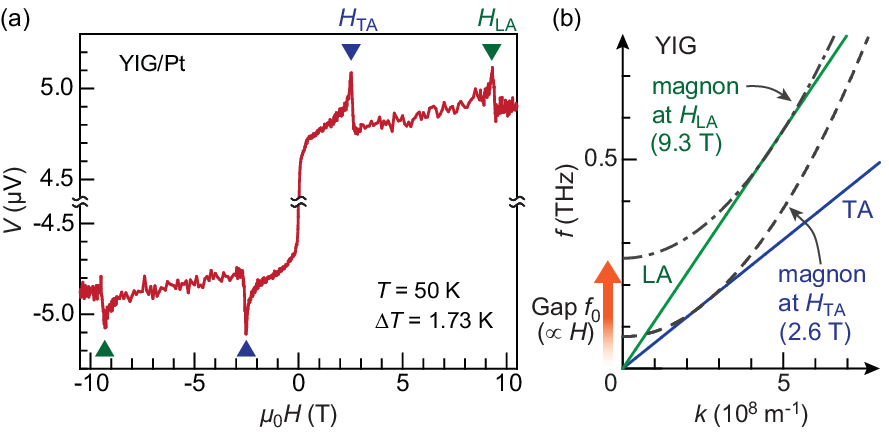}
\caption{(a) $H$ dependence of longitudinal SSE voltage, $V$, for a YIG-film/Pt system at the temperature $T=50~\textrm{K}$ under the temperature difference $\Delta T=1.73~\textrm{K}$ \cite{Kikkawa2016PRL}. 
(b) Magnon, TA-, and LA-phonon dispersion relations for YIG at the touching fields, $H=H_{\rm TA}$ ($=2.6~\textrm{T}$) and $H_{\rm LA}$ ($=9.3~\textrm{T}$). 
The magnon gap, $f_0$, can be tuned with the external $H$ due to the Zeeman interaction ($f_0 \propto \gamma \mu_0 H$). (a) Reproduced from Kikkawa {\it et al.}, Phys. Rev. Lett. {\bf 117}, 207203 (2016). Copyright 2016 American Physical Society.
} \label{fig:MP-SSE}
\end{center}
\end{figure}
\begin{figure*}[htb]
\begin{center}
\includegraphics[width=15.5cm]{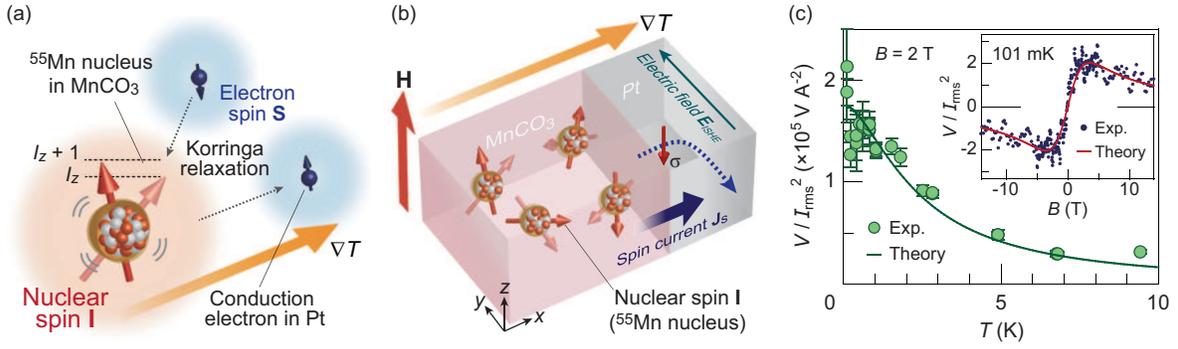}
\end{center}
\caption{(a) Illustration of the nuclear SSE induced by the Korringa relaxation process, the spin-conserving flip-flop scattering between a nuclear spin, {\bf I}, of $^{55}$Mn in MnCO$_3$ and an electron spin, {\bf S}, in Pt via the interfacial hyperfine interaction. 
(b) A schematic illustration of the nuclear SSE in a MnCO$_3$/Pt structure.
(c) $T$ dependence of the nuclear SSE voltage, $V$, (normalized by the applied heat power $\propto I_{\rm rms}^2$) at $B = 2~\textrm{T}$. The inset shows the $B$ dependence of $V/I_{\rm rms}^2$  at  $T = 101~\textrm{mK}$. Theoretical results for the nuclear SSE are also plotted with solid curves. 
(a)-(c) Reproduced from Kikkawa {\it et al.}, Nat. Commun. {\bf 12}, 4356 (2021). Copyright 2021 Author(s), licensed under a Creative Commons Attribution (CC BY) license.
} \label{fig:NSSE}
\end{figure*}
Recent experiments have revealed that small saw-tooth peak structures appear in magnetic field-dependent SSE voltages  [see Fig. \ref{fig:MP-SSE}(a) that shows the $H$ dependence of the SSE for a YIG/Pt bilayer \cite{Kikkawa2016PRL}], which is explained in terms of excitations of magnon polarons having both magnonic and phononic constituents \cite{Kikkawa2016PRL,Flebus2017PRB}. 
The SSE anomalies appear when the magnon dispersion shifts upward with external $H$, such that the phonon dispersion curves become tangential  [Fig. \ref{fig:MP-SSE}(b)]. 
Under these ``touching'' conditions, the magnon and phonon modes can be coupled over the largest frequency window in momentum space, so the effect of magnon-polaron formation in magnonic spin transport is maximal. 
%
%
If the phonon lifetime is longer than magnons, magnon polarons will have a longer lifetime than pure magnons, and can, thus, enhance the SSE \cite{Kikkawa2016PRL,Flebus2017PRB}. Indeed, the SSE anomalies are well reproduced by solutions of the Boltzmann equation for strongly coupled magnon-phonon systems, in which the magnon-phonon lifetime difference is taken into consideration \cite{Kikkawa2016PRL,Flebus2017PRB}. 
So far, the magnon-polaron anomalies in SSEs have been reported in ferrimagnetic YIG \cite{Kikkawa2016PRL,Cornelissen2017PRB,Oyanagi2020AIPAdv,Shi2021PRL_YIG-bulk,Kikkawa2023ARCMP}, Bi$_x$Y$_{3-x}$Fe$_{5}$O$_{12}$ \cite{Kikkawa2022BiYIG}, Fe$_{3}$O$_{4}$ \cite{Xing2020PRB}, NiFe$_{2}$O$_{4}$ \cite{Shan2018APL}, Ni$_{0.65}$Zn$_{0.35}$Al$_{0.8}$Fe$_{1.2}$O$_{4}$ \cite{Wang2018APL}, (partially) compensated ferrimagnetic Lu$_{2}$Bi$_{1}$Fe$_{4}$Ga$_{1}$O$_{12}$ \cite{Ramos2019NatCommun} and Gd$_3$Fe$_5$O$_{12}$  \cite{Yang2021PRB}, and antiferromagnetic Cr$_2$O$_3$ \cite{Li2020PRL_Cr2O3}. 
These works demonstrate the power of SSEs to reveal spectroscopic information about the spin dynamics in various magnetic insulators. \par
The formation of magnon polarons is predicted to affect magnonic spin and thermal conductivities \cite{Flebus2017PRB}. Also, in a non-local configuration, it appears as a Fulde-Ferrell-Larkin-Ovchinnikov (FFLO)-like oscillatory voltage as a function of injector-detector distance \cite{Rameshti2019PRB}, which is yet to be experimentally confirmed. 
Another interesting challenge would be the elucidation of anisotropic magnon-polaron transport recently found through longitudinal and nonlocal SSE measurements \cite{Kikkawa2022BiYIG}.
It is worthwhile to mention that under sufficiently strong magnon-magnon and phonon-phonon scatterings, the coherent magnon-polaron picture may become invalid.   
Schmidt {\it et al.} \cite{Schmidt2018PRB} formulated a Boltzmann transport theory in such a parameter regime and showed that similar anomalies in the SSE manifest through the ``phonon drag'' process at the touching fields.  
%
%
\subsubsection{Quantum-spin Seebeck effect} \label{sec:qSSE}
%
Collective excitations of localized spins are not limited to spin waves, and more exotic excitations are known to exist in quantum spin liquids (QSLs), and have only been discussed in the context of spintronics recently \cite{Hirobe2017NatPhys,Hirobe2018JAP,Chen2021NatCommun}.
In QSLs, localized spins do not exhibit long-range order, but maintain spin correlation due to the quantum fluctuation reinforced by low dimensionality or frustration \cite{Hirobe2018JAP}. A well established example is a one-dimensional (1D) QSL, in which a spin-1/2 chain is coupled via AF interaction and exhibits a gapless elementary excitation, called a ``spinon''.
In 2017, Hirobe {\it et al.} \cite{Hirobe2017NatPhys} demonstrated the spinon SSE in Sr$_2$CuO$_3$ having 1D Cu$^{2+}$ spin ($S=1/2$) with large nearest-neighbor exchange coupling ($\sim 2000~\textrm{K}$). 
The observed spinon SSE is characterized by two features: the non-saturating $H$-response and negative sign. The former is attributed to the robust gapless feature of spinons with large exchange coupling, whereas the latter is mainly to the singlet correlation in spin chains \cite{Hirobe2017NatPhys,Hirobe2018JAP}.  
This demonstration may serve as a bridge between the spintronics and quantum-spin communities, which have developed independently for many years.  
%
%
Recently, Chen {\it et al.} \cite{Chen2021NatCommun} expanded the concept of the quantum-spin Seebeck effect to a gapped spin system through the demonstration of triplon SSE in CuGeO$_3$.  
Subsequently, Xing {\it et al.} \cite{Xing2022APL} reported the SSE in a spin-gapped quantum magnet, Pb$_2$V$_3$O$_9$, and found a peak behavior at around the critical field, $B_{\rm c}$, for the Bose--Einstein condensation states of triplons. 
These works show that the SSE can be a probe for spin excitations in gapped spin systems, and therefore be applied to other materials with exotic spin excitations, such as spin ladder systems and Shastry-Sutherland systems.  
%
\subsubsection{Nuclear-spin Seebeck effect} \label{sec:NSSE}
%
Until recently, all of the SSEs have been an exclusive feature of electron spins or orbitals, so they inevitably disappear at ultralow temperatures or high fields due to entropy quenching \cite{Kikkawa2015PRB,Kikkawa2016JPSJ}. In a solid, there is an unexplored spin and entropy carrier that is well activated even in such an environment: a nuclear spin.
Because of its tiny gyromagnetic ratio, $\gamma_{\rm n}$ ($\sim 10^3$ times less than that of electrons $\gamma$), a nuclear spin exhibits much lower excitation energy than that of electron spins in ambient fields \cite{Rezende2022JAP}, allowing its thermal agitation.  
In 2021, Kikkawa {\it et al.} \cite{Kikkawa2021NatCommun} reported an observation of the nuclear-spin Seebeck effect. 
The material of choice is MnCO$_3$, having a large nuclear spin ($I = 5/2$) of $^{55}$Mn nuclei and strong hyperfine coupling \cite{Shiomi2019NatPhys}, with a Pt contact.  
The observed nuclear SSE is enhanced down to 100 mK and is not suppressed, even under the strong field 14 T, distinct from the electronic SSEs [Fig. \ref{fig:NSSE}(c)]. The voltage features are attributed to entropic nuclear-spin excitation with a tiny energy scale of $\sim 30~\textrm{mK}$, which is minimally affected by the field. The result is quantitatively reproduced by a nuclear SSE theory in which interfacial Korringa process \cite{Korringa1950Physica} is taken into consideration \cite{Kikkawa2021NatCommun} [see Figs. \ref{fig:NSSE}(a) and \ref{fig:NSSE}(c)].
The work may serve as the bridge between nuclear-spin science and thermoelectricity.  
The spin-current mechanism based on the Korringa relaxation may be important to find other nuclear spintronic phenomena. \par 
%
%
%
%
\subsection{Spin Peltier effect} \label{sec:SPE}
Onsager's reciprocity relation indicates the existence of the Onsager equivalent of SSE: spin Peltier effect (SPE) referring to the heat-current generation as a result of a spin current in a metal/magnet system \cite{Flipse2014PRL}.  
In the SPE, a charge current, ${\bf J}_{\rm c}$, applied to a metal induces a spin accumulation at the interface with the magnet due to the SHE \cite{Hoffman2013IEEE,Sinova2015RMP}, which creates or annihilates a magnon in the magnet via the interfacial spin-exchange interaction [see Fig. \ref{fig:SSE-SPE}(b)]. 
Because of energy conservation, this process accompanies a heat (energy) transfer between the electron in the metal and the magnon in the magnet, leading to a temperature difference between these systems. 
Experimentally, SPE-induced temperature modulation has been detected with a thermocouple \cite{Flipse2014PRL,Itoh2017PRB,Yahiro2020PRB}, lock-in thermography (LIT)  \cite{Daimon2016NatCommun,Daimon2017PRB,Uchida2017PRB,Seki2018APL,Yagmur2018JPhysD,Daimon2020APEX}, lock-in thermoreflectance \cite{Yamazaki2020PRB}, and heat-flux sensor based on Peltier cells \cite{Sola2019SciRep}.  
The reciprocal relation between the SPE and SSE is addressed both experimentally and theoretically in Refs. \onlinecite{Sola2019SciRep,Ohnuma2017PRB,Basso2018IEEE,Daimon2020APEX}. \par
So far, a few studies address the SPE at low temperatures \cite{Yagmur2018JPhysD,Yahiro2020PRB}. By means of LIT, Yagmur {\it et al.} \cite{Yagmur2018JPhysD} measured the $T$ dependence of the SPE in a Pt/Gd$_3$Fe$_5$O$_{12}$ system from 300 to 281 K and observed a sign change at around the magnetic compensation temperature, $T_{\rm comp} = 288~\textrm{K}$, for Gd$_3$Fe$_5$O$_{12}$. The SPE in this system is expected to show another sign change at a lower $T$ of $\sim 71~\textrm{K}$ due to the competition of  multiple magnon modes. However, it is not easily accessible through the LIT method, as the output infrared emission intensity ($\propto T^4$) is very weak over such a temperature range \cite{Yagmur2018JPhysD}.  
Yahiro {\it et al.} \cite{Yahiro2020PRB} investigated the magnon-polaron SPE in a Pt/Lu$_{2}$Bi$_{1}$Fe$_{4}$Ga$_{1}$O$_{12}$ system down to 100 K using a thermocouple sensor, but found it difficult to measure further low-$T$ as the thermocouple becomes less sensitive with decreasing $T$. Different types of thermometry would be required to further investigate the SPE at low temperatures, for a wide range of materials. 
 
%
%
\section{\label{Spin-mechanical coupling}Spin-mechanical coupling}
\begin{figure}[htbp]
\begin{center}
\includegraphics[width=7cm, clip]{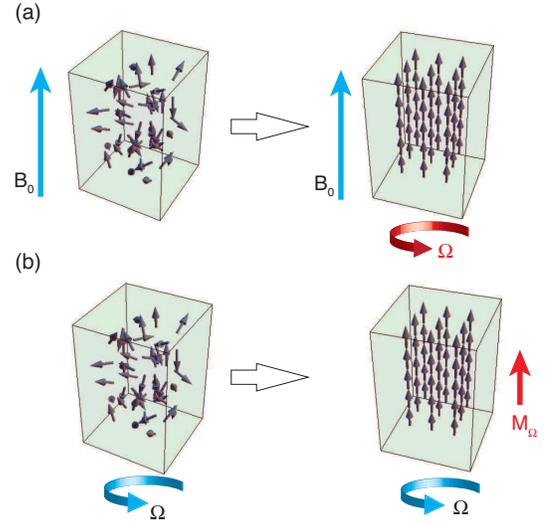}
\end{center}
\caption{
Illustrations of (a) the Einstein-de Haas effect and (b) the Barnett effect.
}
\label{fig:Barnett}
\end{figure} 

\begin{figure}[htbp]
\begin{center}
\includegraphics[width=9cm, clip]{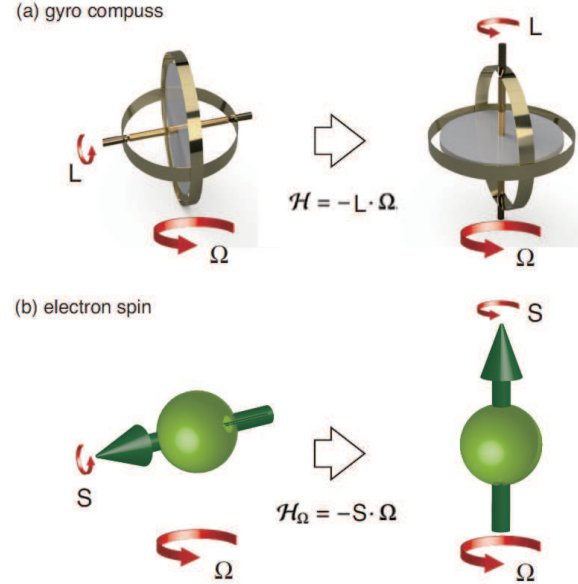}
\end{center}
\caption{
Schematic illustrations of (a) a gyro compass and (b) spin-rotation coupling.
}
\label{fig:gyro}
\end{figure} 
Spin mechanical coupling played a crucial role in developing quantum mechanics.
The coupling between electron spin and mechanical rotation provided the first experimental proof that an electron has an angular momentum \emph{i.e.}, spin, which was reported by Einstein, de Haas, and Barnett in 1915\cite{Barnett1915, Barnett1935, Einstein1915, Scott1962}.
Nowadays, these effects are known as the Einstein-de Haas (EdH) effect and the Barnett effect.
The EdH effect is the phenomenon, in which a magnetic material is magnetized by applying a magnetic field, such that the spin angular momentum is aligned, resulting in the material mechanically rotated due to the angular momentum conservation law, as shown in Fig. \ref{fig:Barnett}(a)\cite{Einstein1915, Scott1962}.
The reverse of the EdH effect is referred to as the Barnett effect, in which a mechanically rotating magnet is magnetized as shown in Fig. \ref{fig:Barnett}(b)\cite{Barnett1915, Barnett1935}.
By using these results, they experimentally determined the value of the $g$ factor of an electron to be $\sim$2 prior to the establishment of modern quantum physics.
The EdH effect has been exploited to determine the $g$ factors of electrons and the orbital component of the magnetic moment in various materials\cite{HUGUENIN1966, HUGUENIN1971}.

The classical analog of the Barnett effect is the Coriolis force acting on the angular momentum in a rotating reference frame\cite{Landau_Mech}.
As shown in Fig. \ref{fig:gyro}(a), when a rotating matter such as a spinning gyrotop, which has an angular momentum, is externally rotated, namely in the rotating reference frame, the rotation axis of the gyrotop aligns parallel to the external rotation axis.
The Hamiltonian of this phenomenon is expressed as
\begin{equation}
	\mathcal{H}=-{\bf L} \cdot {\bm \Omega},
\end{equation}
where ${\bf L}$ is the angular momentum of the spinning gyrotop and ${\bm \Omega}$ is the angular velocity of the external rotation.
The application of this phenomena is a gyro compass, which has been used in vessel navigation since it was invented in 1885.
In the case of a gyro compass, the external rotation is the rotation of the earth, and the rotation axis of the gyro compass points along the rotation axis of the earth.

From the perspective of quantum dynamics, quantum spin angular momentum, ${\bf S}$, also couples to external rotation.
The Hamiltonian is expressed as
\begin{equation}
	\mathcal{H}_{\Omega}=-{\bf S} \cdot {\bm \Omega}.
\label{eq:SR}
\end{equation}
This coupling is referred to as spin-rotation coupling, and is rigorously derived from general relativistic quantum theory\cite{Hehl1976, Hehl1990}.
The spins and associated magnetic moments in the rotating material align parallel to the direction of the rotation axis.
%
%
As a result, the rotating material is magnetized.
This is the mechanism of the Barnett effect.
As shown in Eq. (\ref{eq:SR}), the external rotation couples with the spin angular momentum similarly to how the magnetic field, ${\bf B}$, couples with the magnetic moment, which is referred to as the Zeeman interaction, $\mathcal{H}_Z=-{\bf M} \cdot {\bf B}$.
As the Zeeman interaction is the most fundamental coupling for measuring magnetization, we might be able to say that the Barnett effect is the most fundamental phenomenon for measuring spin angular momentum in a material.

Along with latest advances in physics and technology, the coupling between a spin and mechanics has once again attracted attention, especially in terms of spintronics.
The EdH effect has been exploited to mechanically manipulate micro devices using a spin angular momentum, such as cantilever and paddles \cite{Harii2019, Zolfagharkhani2008}.
The Barnett effect has also been exploited to generate a spin current from mechanical motion, such as fluid flows and surface acoustic waves \cite{Takahashi2015, Takahashi2020, Kobayashi2017}.

\subsection{\label{sec:Barnett effects}Barnett effects}
The Barnett effect has only been studied in ferromagnets with large magnetization at room temperature prior to our studies.
However, considering that the origin of the Barnett effect is the spin-rotational coupling acting on a spin of a single particle, the Barnett effect should be observed for not only paramagnetic materials, but also nuclear spin systems.
In this section, we first show the experimental results of the observation of the Barnett field, $B_{\Omega}$, acting on nuclei by using the NMR and the nuclear quadrupole resonance (NQR) methods\cite{Chudo2014, Chudo2015, Harii2015, Chudo2021}.
\subsubsection{Barnett field observed by NMR and NQR}
The $B_{\Omega}$ is the inertial magnetic field acting on a particle possessing finite gyromagnetic ratio, $\gamma$, in the rotating reference frame.
The $B_{\Omega}$ can be derived from Eq. (\ref{eq:SR}) as follows,
\begin{equation}
	\mathcal{H}_{\Omega}=-S \cdot \Omega = -\gamma S \cdot B_{\Omega},
\end{equation}
where $S$ and $\Omega $ are the magnitude of ${\bf S}$ and ${\bm \Omega}$, respectively, and $B_{\Omega}=\Omega / \gamma$.

%
Here, in the field of magnetic resonance methods, such as ESR and NMR, $\Omega / \gamma$ is the same form as the so-called fictitious or ghost field, $\omega_{\rm L}/\gamma$, which emerges in the rotating reference frame with the Larmor frequency, $\omega_{\rm L}$.
In standard textbooks of magnetic resonance methods, to simplify the spin dynamics exited by the rf field or microwave under a static external magnetic field, classical Newtonian rotational coordinate transformation with the angular velocity of $\omega_{\rm L}$ is introduced \cite{Abragam, Slichter}.
Due to these misleading name, such as 'fictitious' or 'ghost' fields, it has been often misinterpreted that these fields as a non-real field, which means that these fields do not lift the energy level of spin states split by the Zeeman interaction due to the external field and do not generate magnetization\cite{Arabgol2019, Jeener2020}.
Nevertheless, the fictitious or ghost field described as $\omega_{\rm L}/\gamma$ can cancel the existent external magnetic field in the rotational reference frame.
Therefore, the interpretation of the fictitious or ghost field has been ambiguous in the framework of the Newtonian mechanics. 
On the basis of the general gauge theory, however, $\omega_{\rm L}/\gamma$ emerging in the rotational reference frame is the inertial electromagnetic field caused by the rotational coordinate transformation.
Furthermore, the inertial field cannot be locally distinguishable from a real field \cite{Hehl1976, Hehl1990, Frohlich1993, Chudo2020}. 
This fact is guaranteed by the equivalence principle.
%
%
Then, we would like to mention our interpretation on why $\omega_{\rm L}/\gamma$ is called the fictitious or ghost field.
The reason is very simple, i.e., a sample is stationary in the usual magnetic resonance.
Although a sample is stationary, the rotational coordinate transformation is introduced and considered $\omega_{\rm L}/\gamma$ so as to simplify the spin dynamics.
Therefore, in reality, $\omega_{\rm L}/\gamma$ does not act on the stationary sample.
Thus, generally, $\omega_{\rm L}/\gamma$ induced by the rotational coordinate transformation is called the fictitious or ghost field.
%
%
However, when the sample is actually rotating, $\Omega/\gamma$ acts as a real field on a spin of particle possessing a finite $\gamma$ value in the rotating sample.
With this viewpoint, the Barnett effect, in which mechanical rotation induces magnetization, can be simply described to be $M_{\Omega}=\chi B_{\Omega}$ , where $M_{\Omega}$ is a magnetization induced by the mechanical rotation and $\chi$ is a magnetic susceptibility of the sample\cite{Landau, Hahn, Frohlich1993, Chudo2020}.
Hereafter, we demonstrate the observation of the Barnett field, $B_{\Omega}$, in the rotating sample by NMR and NQR methods.

\begin{figure}[tbp]
\begin{center}
\includegraphics[width=8cm, clip]{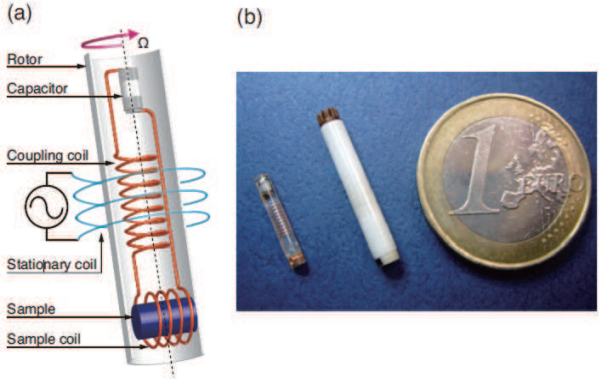}
\end{center}
\caption{
(a) Illustration of the NMR (NQR) tuning circuit capable of high-speed rotation.
(b) NMR (NQR) tuning circuit and the high-speed rotor.
}
\label{fig:rotor}
\end{figure}

To observe the Barnett field, the signal detector, which is a pickup coil in the case of NMR and NQR, must be in the same rotating reference frame as the rotating sample because the Barnett field is an inertial field that emerges on the rotating reference frame.
To overcome this difficulty, we have developed a new NMR (NQR) tuning circuit with the capacity for high-speed rotation, as shown in Fig. \ref{fig:rotor}(a).
The rf field generated at the NMR spectrometer is transmitted into the inner tuning circuit through the mutual induction between the stationary coil and the coupling coil.
These two coils are electromagnetically coupled, but mechanically decoupled.
Thus, we can keep electromagnetic coupling during high-speed rotation of the inner circuit.
The key technology of this setup is the wireless connection between a rotating frame and a laboratory frame, which enables us to observe the spin dynamics on the rotating reference frame.
In Fig. \ref{fig:rotor}(b), we show the tuning circuit and the high-speed rotor used in the study described here\cite{Chudo2014}.
The tuning circuit is embedded into epoxy resin to prevent centrifugal damage due to the high rotational speed.
The sample is put into the sample coil, and both the sample and the circuit are put into this high-speed rotor.
Then, we blow compressed air into the air turbine to rotate the rotor.
This technique enables us to rotate only the sample coil with the sample fixed in the stationary laboratory reference frame.
%
Using these techniques, we systematically study the effects of rotation in the setups involving only the sample coil rotation, only the sample rotation, and simultaneous sample coil and sample rotation.
Applying these setups to NMR and NQR measurements, we observe the NMR line shifts and NQR line splittings, in which the spectral structures are clearly distinct.

\begin{figure}[tbp]
\begin{center}
\includegraphics[width=8cm, clip]{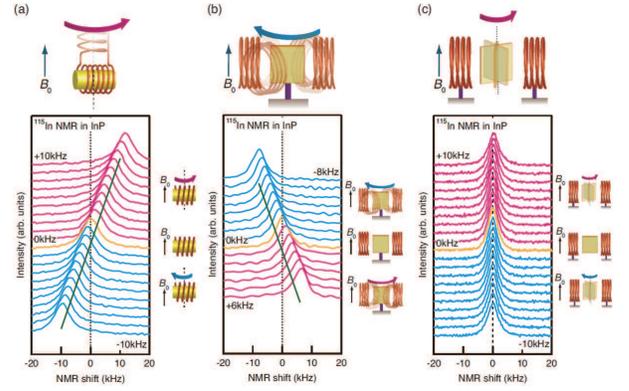}
\end{center}
\caption{
$^{115}$In NMR spectra in InP obtained by using the setup with (a) simultaneous sample coil and sample rotation, (b) only the sample coil rotation, and (c) only the sample rotation.
}
\label{fig:JPSJ}
\end{figure}

Figure \ref{fig:JPSJ}(a) shows the $^{115}$In NMR in InP using the setup shown in Fig. \ref{fig:rotor}(a) under the external magnetic field parallel to the rotation axis.
With positive rotation, the NMR shift proportionally increases by increasing the rotation frequency. By contrast, with negative rotation, the NMR shift proportionally decreases by decreasing the rotation frequency.
The value of the NMR shift, $\Delta f$, coincides with the rotation frequency, $\Omega/2\pi$, \emph{i.e.}, $2\pi \Delta f=\gamma B_{\Omega}=\Omega$.
This behavior is completely consistent with the Barnett field, $B_{\Omega}$, acting on the nuclei in the rotating sample.

Using the setup in Fig. \ref{fig:rotor}(a), we have been able to conduct NMR measurement in the rotating reference frame.
Then, we customized this setup to rotate only the sample coil with the sample fixed in the laboratory reference frame.
In this way, we can observe the effect of the relative rotational motion between the sample coil and sample using NMR.
The results are shown in Fig. \ref{fig:JPSJ}(b).
The NMR shift also occurs according to the relative rotational motion.
The value of the NMR shifts coincides with the rotation frequency, $\Omega/2\pi$.
Thus, we call this NMR shift the rotational Doppler effect.
It should be noted here that, even though the value of the NMR shifts arising from the Barnett effect and the rotational Doppler effect is equivalent, their origins are different.
In the former case, there is no relative rotation between the sample coil and sample, therefore, the NMR shift arises from the inertial magnetic field, so-called the Barnett field.
The latter arises from the relative rotational motion between the sample coil and sample because the sample is in the laboratory reference frame.
Figure \ref{fig:JPSJ}(c) shows the NMR spectra with the setup of only the sample rotation.
In this case, there is no NMR shift.
The reason for this is that both the Barnett field and rotational Doppler effect are present.
In the case of sample rotation, the direction of the relative rotation is opposite to the rotational direction of the sample.
Therefore, these effects cancel each other so that there is no NMR shift because the absolute values of the NMR shifts in both effects are equivalent.

\begin{figure}[htbp]
\begin{center}
\includegraphics[width=8cm, clip]{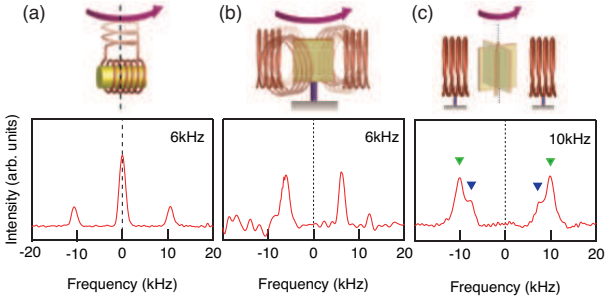}
\end{center}
\caption{
$^{35}$Cl NQR spectra in NaClO$_3$ obtained by using the setup of (a) simultaneous sample coil and sample rotation, (b) only sample coil rotation, and (c) only sample rotation.
}
\label{fig:NQR}
\end{figure}

However, there is a criticism that both NMR shifts shown in Figs. \ref{fig:JPSJ}(a) and (b) have the same origin, \emph{i.e.}, sample rotation is irrelevant and, thus, coil rotation causes the NMR shift and the Barnett field is an unnecessary hypothesis\cite{Jeener2020}.
In this interpretation, it can be only thought that nuclear spins inside a rotating sample isolate from the rotational motion of the sample.
However, as we claim in Sec. \ref{Spin-mechanical coupling}, the nuclear spin in the rotating sample is in the rotating reference frame and, thus, the spin rotation coupling acts on the nuclear spin.
Therefore, the Barnett field as the inertial magnetic field in the rotational reference frame acts on the nuclear spins.
The interpretation that nuclear spins in a rotating sample isolate from the rotating sample is irrelevant. 
In the next paragraph, we provide the experimental evidence that nuclear spin in the rotating sample couples with the sample rotational motion\cite{Chudo2021}.

Figure \ref{fig:NQR} shows $^{35}$Cl NQR spectra in the single crystal of NaClO$_3$ obtained by the setup of (a) the simultaneous sample and sample coil rotation, (b) only the sample coil rotation, (c) only the sample rotation.
The rotation axes are parallel to the <100> direction of the single crystal.
In the case of the simultaneous sample and sample coil rotation, the NQR spectra split into three lines with the NQR shifts of $\pm \sqrt{3} \Omega /2 \pi$.
This spectral structure can be reproduced by treating the Barnett field as a perturbation to the primal quadrupole Hamiltonian.
This result is explicit proof of the existence of the Barnett field in the rotating reference frame.
By contrast, in the case of only the sample coil rotation, the NQR spectra split into two lines with NQR shifts of $\pm \Omega /2 \pi$.
This NQR line splitting is caused by the relative rotational motion between the sample and sample coil, namely, the rotational Doppler effect.
Compared with these two NQR spectra, it is obvious that the simultaneous sample and sample coil rotation exhibits unique phenomena from the only the sample coil rotation.
In the case of just sample rotation shown in Fig. \ref{fig:NQR}(c), a different NQR spectral structure is also observed when compared to Fig. \ref{fig:NQR}(a) and (b), meaning that the NQR shift due to the Barnett field and the rotational Doppler effect do not cancel each other out.
This result is critically different from the NMR results shown in Fig. \ref{fig:JPSJ}(c).
For more details about the analysis of the NQR spectra, please refer to Ref. \onlinecite{Chudo2021}.

\subsubsection{Barnett effect on paramagnet}
We demonstrate the experimental results of the Barnett effect in electronic spin systems of paramagnetic state\cite{Ono2015}.
To observe the Barnett effect on the paramagnetic state, we first targeted Gd as a sample because of its large magnetic moment arising from the seven 4\textit{f} electrons.
In addition, Gd shows ferromagnetic transition near room temperature (292 K).
Therefore, at just above 292 K, Gd has a large magnetization, even in the paramagnetic state.
By improving the accuracy of the apparatus, e.g. stabilizing the temperature and strengthening the magnetic shield, we succeeded in observing the Barnett effect on Tb and Dy\cite{Ono2015, Ogata2017rare}.
4\textit{f} electrons are positioned in the inner core inside ions, and have a local magnetic moment isolated from the lattice system.
This situation is similar to the nuclear spin system.
Therefore, as the Barnett effect of the 4\textit{f} electron system can be observed, the 4\textit{f} electron in the rotating reference frame feels the Barnett field as an inertial magnetic field in the rotating reference frame.

\begin{figure}[tbp]
\begin{center}
\includegraphics[width=8cm, clip]{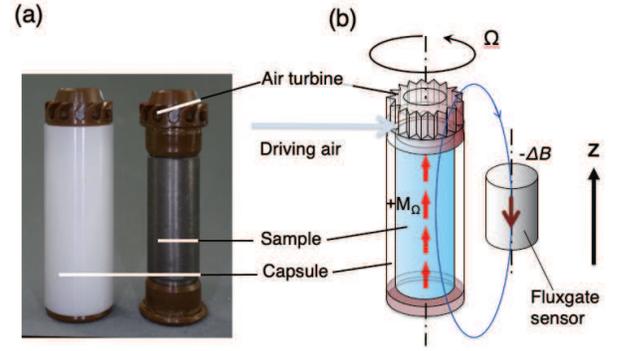}
\end{center}
\caption{
Experimental setup for observation of the Barnett effect. (a) High-speed rotor (left) and sample (right). (b) Illustration of the setup. The directions of magnetization, stray field detected by fluxgate sensor, and rotation are defined as indicated by arrows.
}
\label{fig:setupBarnett}
\end{figure}

Figure \ref{fig:setupBarnett} shows the experimental setup for observing the Barnett effect.
The cylindrically shaped sample is inserted into the high-speed rotor.
The rotor is installed into the high-speed rotation system, which consists of an air bearing and two driving air channels.
The rotation system is originally produced by JEOL for magic angle spinning NMR measurements and is improved to realize the two-way rotation, that is, backward and forward directions.
High-speed rotation is realized by blowing compressed air into the air turbine attached at the sample tube.
By switching the direction of air flow, the rotational direction can be reversed.
The fluxgate magnetic sensor mounted adjacent to the rotation system measures the stray field $\Delta B$ from the sample magnetized by the Barnett effect.
The definition of the directions of magnetization, the stray field, and rotation are shown in Fig. \ref{fig:setupBarnett}(b).
The rotation system and the fluxgate magnetic sensor are enclosed in the magnetic shield made of permalloy.
The magnetic shield is composed of two layers.
To stabilize the temperature, the whole apparatus is placed inside a thermal isolation chamber, where the temperature is controlled within $\pm$0.1 K using a high precision air controller.
All measurements were performed at room temperature.
The magnetization by the Barnett effect, $M_{\Omega}$, is estimated from $\Delta B$ by using the dipole model.
The Barnett field, $B_{\Omega}$, is estimated from $M_{\Omega}/\chi$.

\begin{figure}[htbp]
\begin{center}
\includegraphics[width=8cm, clip]{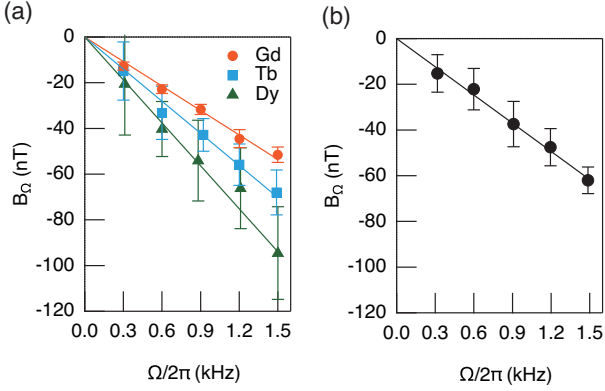}
\end{center}
\caption{
Barnett field of (a) Gd, Tb, and Dy, and (b) FeCo nanogranule, respectively.
}
\label{fig:rareBarnett}
\end{figure}

Figure \ref{fig:rareBarnett}(a) shows the rotation frequency dependence of the Barnett field.
The negative slopes of the data indicate the antiparallel coupling between the spin and the magnetic moment due to the negative charge of an electron.
The gyroscopic $g$-factor, $g'$, can be estimated from the Barnett field using the equation $B_{\Omega}=\frac{2m_{\rm e}}{g'e}\Omega$. 
%
The $g'$ factors are estimated to be 2.00$\pm$0.08, 1.53$\pm$0.17, and 1.15$\pm$0.32 for Gd, Tb, and Dy, respectively\cite{Ogata2017rare}.
These values are very close to the Lande's $g$ factor, which is 2, 3/2, and 4/3 for Gd, Tb, and Dy, respectively.
This fact indicates that the 4\textit{f} electron states are well described by the LS coupling scheme.

In the case of 3\textit{d} electron systems, the magnetic moment mainly arises from the spin component.
The gyroscopic $g'$ factor can be used to estimate the orbital contribution to the magnetic moment.
To demonstrate this, we measured the Barnett effect on the super paramagnet, which has the large magnetization comparable to the ferromagnet and behaves paramagnetically \cite{Ogata2017nano}.
The sample is the FeCo nanogranules embedded in a matrix of MgF$_2$\cite{Masumoto2014}.
Figure \ref{fig:rareBarnett}(b) shows the rotation speed dependence of the Barnett field.
From the data slope, the gyroscopic $g'$ factor is estimated to be 1.76$\pm$0.11 \cite{Ogata2017nano}.
When the magnetic moment completely arises from the spin component, $g'$ becomes 2, as is the case with Gd, which has no orbital moment.
However, when the orbital component contributes to the magnetic moment, $g'$ becomes smaller than 2.
Although the $g'$ factor of the bulk FeCo is 1.916, the $g'$ factor of FeCo nanogranules is smaller than the bulk\cite{RECK1969}.
This result indicates that the orbital contribution to the magnetic moment in FeCo nanogranules is larger than the bulk because orbital magnetism is enhanced due to symmetry breaking of the electron system at the surface of the nanogranules, and the fraction of the surface of FeCo nanogranules is larger than the bulk\cite{Wu1992}.

\subsection{Angular momentum compensation observed by the Barnett effect}
\begin{figure}[tbp]
\begin{center}
\includegraphics[width=9cm, clip]{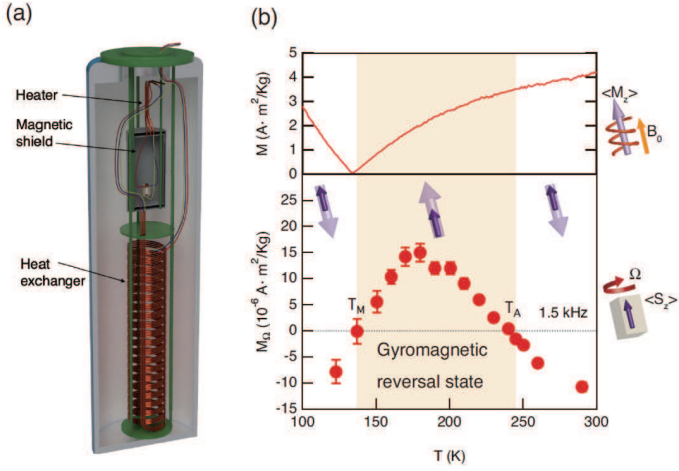}
\end{center}
\caption{
(a) Schematic of the temperature-controllable equipment for observation of the Barnett effect at low temperatures.
(b)
Temperature dependence of the magnetization induced by the magnetic field (upper panel) and the $M_{\Omega}$ induced by the Barnett effect (lower panel) in Ho$_3$Fe$_5$O$_{12}$.
Light and dark purple arrows represent the net magnetic moment and net angular momentum, respectively.
}
\label{fig:LowBarnett}
\end{figure}

Hereafter, we introduce the new application of the Barnett effect.
The Barnett effect can measure the angular momentum in a material.
Therefore, we challenged it to measure the angular momentum compensation temperature in ferrimagnets using the Barnett effect\cite{Imai2018, Imai2019}.
The angular momentum compensation is the singularity where the net angular momentum in the material vanishes at a certain temperature.
Angular momentum compensation typically exists in an N-type ferrimagnet, which has the magnetic compensation temperature, $T_M$ \cite{Pauthenet1958}.
Ferrimagnets consist of two (or more) sublattice possessing different values of magnetic moments, which couples antiparallel to each other.
When the temperature dependence of the order parameter at two sublattices is different, the magnetization at the two sublattices becomes equivalent at a certain temperature, which is $T_M$.
In a general case, when the $g$ factors of the magnetic moment at each sublattice are different, the temperature at which the angular momentum at each sublattice becomes equivalent may be different from $T_M$.
%
This temperature is the angular momentum compensation temperature, $T_A$.
To measure $T_A$ by the Barnett effect, we developed the temperature controllable equipment ranging from room temperature down to 120 K, as shown in Fig. \ref{fig:LowBarnett}(a)\cite{Imai2018}.
The rotation system is installed in the cryostat and cooled by nitrogen gas evaporated from liquid nitrogen.
The bearing and driving gas were high pressed nitrogen, instead of compressed air, and cooled at the heat exchanger inside the cryostat.
Then, the temperature of the cooled high pressed nitrogen gas is controlled by the heater.
The rotation system is inside the magnetic shield.

Figure \ref{fig:LowBarnett}(b) shows the temperature dependence of the magnetization induced by the magnetic field (upper panel) and the $M_{\Omega}$ induced by the Barnett effect (lower panel) in Ho$_3$Fe$_5$O$_{12}$ (HoIG).
In the upper panel, the magnetization disappears at 135 K.
This temperature is the magnetic compensation temperature, $T_M$.
In the lower panel, the $M_{\Omega}$ also disappears at 135K.
This is a trivial behavior because even though the spin responds to the rotation and aligns to the rotation axis due to the Barnett effect, the magnetization is basically zero.
Thus the $M_{\Omega}$ also becomes zero.
Interestingly, the $M_{\Omega}$ also disappears at 240 K despite the finite spontaneous magnetization at 240 K, as shown in the upper panel in Fig. \ref{fig:LowBarnett}(b).
This temperature is $T_A$.
Although, at $T_A$ the spontaneous magnetization has a finite value, the net angular momentum is compensated, and thus, the Barnett effect does not occur.
$T_A$ is also observed by the NMR measurement in HoIG, in which the NMR intensity is enhanced at $T_A$\cite{Chudo2021Barnett, Imai2020}.
At temperatures between 135  and 240 K, $M_{\Omega}$ assumes positive values, whereas, above 240 K and below 135 K, it assumes negative values.
As shown in Figs. \ref{fig:rareBarnett}(a) and \ref{fig:rareBarnett}(b) the $M_{\Omega}=\chi B_{\Omega}$ usually is negative due to antiparallel coupling between a spin and a magnetic moment.
Thus, the positive values of the $M_{\Omega}$ indicate the parallel coupling between a spin and a magnetic moment.
We call this reversal of the coupling the gyromagnetic reversal state.

\begin{figure}[tbp]
\begin{center}
\includegraphics[width=8cm, clip]{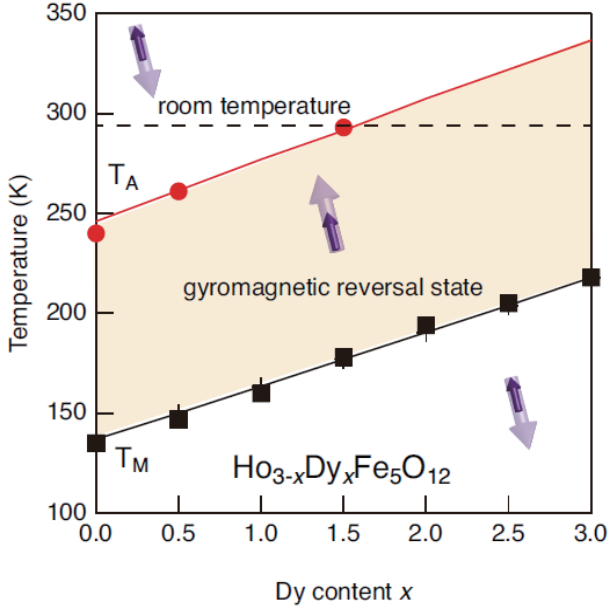}
\end{center}
\caption{
Gyromagnetic phase diagram of Ho$_{3-x}$Dy$_x$Fe$_5$O$_{12}$.
Black squares and red circles represent $T_M$ and $T_A$, respectively.
The beige area represents the gyromagnetic reversal state.
Light and dark purple arrows represent the net magnetic moment and net angular momentum, respectively.
}
\label{fig:PhaseBarnett}
\end{figure}

Angular momentum compensation is thought to be a promising candidate to quicken the speed of magnetization switching, which is important for speeding up a magnetic device using the direction of magnetization as an information carrier because, at $T_A$, there is no inertia due to vanishing net angular momentum in materials.
Thus, manipulation of the angular compensation temperature to room temperature is an important subject for future high-speed magnetic devices.
Recently, we demonstrated the manipulation of $T_A$ in HoIG as a platform by partially substituting Dy for Ho.
The $T_A$ is determined by the Barnett effect.
The results are shown in Fig. \ref{fig:PhaseBarnett}\cite{Chudo2021Barnett, Imai2019}.
Both $T_M$ and $T_A$ increase with increasing Dy content.
The $T_A$ value coincides with room temperature (293 K) at the composition of Ho$_{1.5}$Dy$_{1.5}$Fe$_5$O$_{12}$.
Using the Barnett effect, we can conveniently determine $T_A$ values for any sample, irrespective of its electronic conductivity and state (single crystals, powders, and poly-crystals).
Thus, our studies pave a new way to explore new materials possessing $T_A$, which will be utilized for magnetic devices in the future.

%
%
\section{Conclusions}
In this article, we gave an overview on current studies of spin, spin current, and their related phenomena based on the authors' expertise. The topics include spin pumping, spin Seebeck and Peltier effects, spin transfer and topological Hall torques, emergent inductor, and spin-mechanical/lattice coupling phenomena, where both the electron and nuclear spins and their angular-momentum conversion play an important role. \par

Nowadays, the concept of spin current has emerged in various areas of condensed matter physics and has played a role as a useful guiding principle to open up new phenomena in condensed matter. Moreover, the concept of spin current appears not only in physics, but also in various fields of science and technology. The diversity of materials and device structures has expanded spin-current application possibilities, and cross-disciplinary expansion is an important aspect of this field.
Another interesting physical aspect of spin currents is their universality, which explains a wide range of phenomena in a unified manner with a small number of principles, and their power to predict unknown physical phenomena based on this universality. We anticipate that spin current and its versatile coupling with other physical entities may play an essential role in future electronic devices and technologies, including quantum information science and energy harvesting.

\begin{acknowledgments}
The authors thank T. Makiuchi, T. Hioki, H. Arisawa, Y. Yamane, Y. Araki, M. Matsuo, M. Imai, K. Harii, Y. Ogata, M. Sato, and Y. Ohnuma for valuable discussions, Y. Haga for the X-ray diffraction experiment support and M. Ono for technical support.
This work was financially supported by JST ERATO ''Spin Quantum Rectification Project'' (Grant No. JP- MJER1402), JST CREST (Grants Nos. JPMJCR19J4, JPMJCR1874, JPMJCR20C1 and JPMJCR20T2), JSPS KAKENHI (Grants Nos. JP19H05600, JP19H05622, JP17H02927, JP20H01863, JP20H01865, JP20H02599, JP21H01800, JP21H04643, and JP22K18686), Grant-in-Aid for Transformative Research Areas (No. JP22H05114), Institute for AI and Beyond of the University of Tokyo, IBM-UTokyo lab, and Daikin Industries, Ltd. 

\end{acknowledgments}

\section*{AUTHOR DECLARATIONS}
\subsection*{Conflict of Interest}
\noindent
The authors have no conflicts to disclose.

\subsection*{Author Contributions}
\noindent
{\bf Sadamichi Maekawa:} Conceptualization (equal); Writing – original draft (equal); Writing – review \& editing (equal).
{\bf Takashi Kikkawa:} Conceptualization (equal); Writing – original draft (equal); Writing – review \& editing (equal).
{\bf Hiroyuki Chudo:} Writing – original draft (equal); Writing – review \& editing (equal).
{\bf Jun'ichi Ieda:} Conceptualization (equal); Writing – original draft (equal); Writing – review \& editing (equal).
{\bf Eiji Saitoh:} Conceptualization (equal); Writing – original draft (equal); Writing – review \& editing (equal).

\section*{DATA AVAILABILITY}
The data that support the findings of this study are available from the corresponding author upon reasonable request.

\bibliographystyle{apsrev4-1.bst}
\bibliography{aipsamp_v4}

\end{document}